\documentclass[11pt]{article}

\usepackage[T1]{fontenc}
\usepackage[utf8]{inputenc}
\usepackage{lmodern}
\usepackage{amsmath,amssymb,amsthm}
\usepackage{graphicx}
\usepackage{float}
\usepackage{bm}
\usepackage{booktabs}
\usepackage{microtype}
\usepackage[margin=1in]{geometry}
\usepackage[numbers,sort&compress]{natbib}
\usepackage{hyperref}
\usepackage[capitalise,noabbrev]{cleveref}
\usepackage{xcolor}
\usepackage{url}

\hypersetup{
    colorlinks=true,
    linkcolor=blue,
    citecolor=blue,
    urlcolor=blue,
    pdftitle={BKT-like Correlation Scaling and Twist Responses in a One-Dimensional Fractional U(1) Ginzburg--Landau Model},
    pdfauthor={Hitomi Endo and Michikazu Kobayashi}
}

\title{BKT-like Correlation Scaling and Twist Responses in a One-Dimensional Fractional $U(1)$ Ginzburg--Landau Model}

\author{
Hitomi Endo\\
School of Engineering Science, Kochi University of Technology\\
\texttt{295092w@gs.kochi-tech.ac.jp}
\and
Michikazu Kobayashi\thanks{Corresponding author: \texttt{kobayashi.michikazu@kochi-tech.ac.jp}}\\
School of Engineering Science, Kochi University of Technology\\
\texttt{kobayashi.michikazu@kochi-tech.ac.jp}
}

\date{}

\begin{document}

\maketitle

\begin{abstract}
We study a one-dimensional fractional $U(1)$ Ginzburg--Landau model whose quadratic part has Fourier multiplier $|k|^\sigma$, focusing on the marginal case $\sigma=1$.
This dispersion yields logarithmic spin-wave fluctuations, suggesting BKT-like behavior despite the one-dimensional setting.
We sample the equilibrium Gibbs measure using stochastic Gross--Pitaevskii dynamics and analyze correlation functions, dimensionless ratios, effective exponents, and twist responses.
The correlation function shows a low-temperature algebraic branch with a temperature-dependent exponent, while the high-temperature regime exhibits a nonlocal-kernel-induced tail consistent with $C(r)\sim r^{-2}$.
The correlation and Binder ratios are nearly size-independent at low temperature and collapse with the BKT-type variable $(T-T_{\rm BKT})(\log L)^2$; finite-size effects set in around $T\simeq0.35\text{--}0.4$, consistent with $T_{\rm BKT}\simeq0.35$.
Unlike the two-dimensional XY model, twist responses do not yield a finite helicity modulus: the ordinary linear-response quantity grows with system size, whereas the cusp twist response scales as $L^{-\eta(T)}$, like the squared zero-mode order parameter.
Thus, the transition is BKT-like in correlation scaling, but lacks a universal helicity-modulus jump.
\end{abstract}



\section{Introduction}

Phase transitions associated with continuous symmetries exhibit a subtle interplay between order, fluctuations, topology, and dimensionality.
In conventional Landau theory, a phase transition is characterized by the spontaneous breaking of a symmetry and the emergence of a nonzero local order parameter.
However, this picture is incomplete in low-dimensional systems.
In particular, the Mermin--Wagner theorem excludes spontaneous breaking of continuous symmetries in one- and two-dimensional systems with sufficiently short-range interactions at finite temperature~\cite{MerminWagner1966,Hohenberg1967}.
Nevertheless, two-dimensional systems with $U(1)$ symmetry can exhibit a remarkable topological transition, now known as the Berezinskii--Kosterlitz--Thouless (BKT) transition, where the low-temperature phase possesses quasi-long-range order rather than true long-range order~\cite{Berezinskii1971,Berezinskii1972,KosterlitzThouless1973,Kosterlitz1974}.

The BKT transition is characterized by the binding and unbinding of vortex--antivortex pairs~\cite{KosterlitzThouless1973,Kosterlitz1974}.
When the temperature $T$ is lower than the transition temperature $T_{\rm BKT}$, vortices are bound and the two-point correlation function $C(r)$ decays algebraically,
\begin{align}
    C(r) \sim r^{-\eta(T)},
\end{align}
with a temperature-dependent exponent $\eta(T)$.
Here and below, we use $C(r)\sim r^{-\eta(T)}$ to indicate algebraic decay with exponent $\eta(T)$, in the sense that
\begin{align}
    \eta(T)
    =
    -\lim_{r\to\infty}
    \frac{\log C(r)}{\log r},
\end{align}
whenever this limit exists.
Above the transition temperature, free vortices proliferate and correlations decay rapidly.
A central feature of the standard two-dimensional XY model is that the transition can be characterized not only through the correlation function but also through the helicity modulus, or superfluid stiffness~\cite{FisherBarberJasnow1973,NelsonKosterlitz1977}.
The universal jump of the helicity modulus at the transition provides one of the most striking signatures of BKT physics~\mbox{\cite{NelsonKosterlitz1977,WeberMinnhagen1988,HaradaKawashima1997}}.
From a mathematical viewpoint, the BKT transition is also exceptional because the temperature-dependent correlation length $\xi(T)$ diverges with an essential~singularity,
\begin{align}
    \log \xi(T)
    =
    \frac{b}{\sqrt{T-T_{\mathrm{BKT}}}}
    +
    O(1)
    \qquad
    (T\downarrow T_{\rm BKT}),
    \label{eq:BKT_correlation_length}
\end{align}
rather than with an ordinary power law.

Beyond its original formulation in the two-dimensional XY model, BKT physics appears in a wide variety of systems, including thin superfluid films, superconducting films and Josephson-junction arrays, two-dimensional Bose gases, exciton-polariton condensates, and ultracold atomic gases~\cite{BishopReppy1978,BeasleyMooijOrlando1979,ResnickGarlandBoydShoemakerNewrock1981,HadzibabicKrugerCheneauBattelierDalibard2006,Roumpos2012}.
In these contexts, the BKT transition provides a universal framework for understanding finite-temperature coherence in systems where true long-range order is suppressed but algebraic order can persist.
The same ideas have also motivated extensive mathematical studies of vortex gases, Coulomb gases, nonlinear field theories, and stochastic equations whose invariant measures are related to Gibbs measures of $U(1)$-symmetric energy functionals.

A natural question is how BKT-type phenomena are modified when the underlying spatial operator is nonlocal.
Nonlocal and fractional differential operators arise in many areas of mathematics and physics, including anomalous diffusion, long-range interacting systems, fractional Schr{\"o}dinger equations, nonlocal Ginzburg--Landau models, and effective descriptions of systems with algebraically decaying interactions~\cite{MetzlerKlafter2000,Laskin2002,FisherMaNickel1972,Sak1973,LuijtenBlote1997}.
A prototypical example is the fractional Laplacian $(-\Delta)^{\sigma/2}$, whose Fourier symbol is $|k|^\sigma$~\cite{CaffarelliSilvestre2007,DiNezzaPalatucciValdinoci2012,Kwasnicki2017}.
Such an operator changes the infrared behavior of the theory and therefore modifies the balance between fluctuations and ordering.
In this sense, one-dimensional systems with fractional spatial operators provide a useful arena in which to ask whether BKT-like quasi-long-range order can emerge outside the standard two-dimensional short-range setting.

Related KT-like mechanisms have also been discussed in higher-dimensional settings when anomalous scaling modifies the long-distance interaction between topological defects.
In compact $(2+1)$-dimensional $U(1)$ gauge theories coupled to matter, anomalous gauge-field scaling can generate logarithmic interactions between monopole defects and lead to a BKT-like deconfinement transition~\cite{KleinertHagenNogueiraFlavioSudboAsle2002,HagenKleinertFlavioNogueiraAsleSudbo2003}.
In the corresponding dual description, the latter problem involves a sine-Gordon-like theory with an anomalous gradient energy proportional to $k^3$, together with renormalization-group flows of stiffness and defect fugacity.
Related numerical studies of logarithmic plasmas have also found nonanalytic transition behavior in both two and three dimensions ~\cite{KragsetSudboNogueira2004}.
These works provide complementary examples showing that BKT-like phenomenology need not be restricted to the conventional two-dimensional short-range setting when the effective long-distance interaction is modified.

Recent studies of two-dimensional long-range XY and Villain models have further highlighted the importance of the interplay between spin-wave fluctuations and topological excitations in determining their critical behavior~\cite{Giachetti2023,WaltherLuisWillsherJosefKnolleJohannes2026}.

In the present work, we study a one-dimensional complex field model with a nonlocal fractional gradient energy.
The equilibrium energy functional is of Ginzburg--Landau type,
\begin{align}
    H[q]
    =
    \sum_k |k|^\sigma |q_k|^2
    +
    g
    \int dx\,
    \left(|q(x)|^2-1\right)^2,
    \label{eq:intro_H}
\end{align}
where $q$ is a complex order parameter and $\sigma$ controls the strength of the nonlocal spatial coupling.
The case $\sigma=1$ is of particular interest because it is marginal in the sense that the spin-wave fluctuation becomes logarithmic, as in the two-dimensional short-range XY model and in related inverse-square long-range models~\cite{Thouless1969,MuraokaTakamotoIdogaki2010,GiachettiDefenuRuffoTrombettoni2021,GiachettiTrombettoniRuffoDefenu2022}.
Thus, although the model is one-dimensional, its long-wavelength fluctuations may display BKT-like behavior.

We investigate this model numerically through stochastic dynamics whose invariant measure is the Gibbs distribution associated with $H[q]$.
In particular, we consider stochastic Gross--Pitaevskii-type dynamics as a sampling method for the equilibrium measure~\cite{Stoof1999,GardinerDavis2003,BlakieBradleyDavisBallaghGardiner2008,CockburnProukakis2012,DeBouardDebusscheFukuizumi2018}.
This formulation is useful for two reasons.
First, it connects the problem to stochastic partial differential equations and nonlocal variational structures.
Second, it provides a natural bridge to physical systems such as superfluids, Bose condensates, and nonlinear Schrödinger equations with damping and noise.
The focus of this paper, however, is on equilibrium static properties rather than on the dynamical universality class.

The main purpose of this work is to clarify whether a one-dimensional fractional $U(1)$ model can exhibit a BKT-like transition and, if so, how such a transition should be characterized.
We analyze the order parameter, Binder ratio, correlation-function ratio, effective correlation exponent, and twist-response quantities analogous to the helicity modulus.
In the standard two-dimensional XY model, the helicity modulus plays a central role as a stiffness that remains finite in the low-temperature phase and exhibits a universal jump at the transition.
In the present nonlocal one-dimensional model, however, the situation is more subtle.
Natural definitions of twist response, including spectral cusp responses, do not behave as direct analogues of the two-dimensional helicity modulus.
Instead, they are strongly tied to the finite-size scaling of the zero-mode order parameter.

Our numerical results indicate that the $\sigma=1$ model exhibits a BKT-like change in the long-distance behavior of the correlation function.
Below the transition temperature, the correlation function displays algebraic decay with a temperature-dependent exponent.
Above the transition, the decay is not simply the ordinary exponential decay familiar from short-range systems.
Rather, the nonlocal fractional kernel induces a long-distance algebraic tail, while a large crossover scale controls the approach to this asymptotic regime.
The finite-size scaling of correlation ratios and Binder ratios is consistent with a BKT-type essential singularity of the corresponding crossover length.
In particular, the scaling is compatible with the standard BKT form in Equation~\eqref{eq:BKT_correlation_length}.

This paper is organized as follows.
In Section~\ref{sec:model}, we define the fractional $U(1)$ model, the associated Gibbs measure, the spectral twist response, and the observables used to characterize the transition.
We also discuss the relation between the Fourier multiplier $|k|^\sigma$ and nonlocal algebraic interactions.
In Section~\ref{sec3}, we describe the stochastic dynamics used to sample the equilibrium Gibbs measure and the numerical procedure.
In Section~\ref{sec4}, we present numerical results for the order parameter, Binder ratio, correlation ratio, effective exponent, and helicity-modulus-like quantities.
In Section~\ref{sec5}, we discuss the interpretation of the observed transition, emphasizing both its similarities to and differences from the standard two-dimensional BKT transition.
Finally, Section~\ref{sec6} summarizes our conclusions and outlines open problems concerning nonlocal $U(1)$ models, fractional Laplacians, and generalized notions of stiffness.

\section{Model and Analytical Considerations}
\label{sec:model}

\subsection{Fractional $U(1)$ Ginzburg--Landau Model}

We consider a one-dimensional complex field $q(x)\in\mathbb C$ on a periodic domain of length $L$.
The equilibrium properties are governed by a nonlocal Ginzburg--Landau energy~\mbox{functional}
\begin{align}
    H[q]
    =
    \int_0^L dx\,
    q^\ast(x)(-\Delta)^{\sigma/2}q(x)
    +
    g
    \int_0^L dx\,
    \left(|q(x)|^2-1\right)^2,
    \label{eq:H_continuum}
\end{align}
where $0<\sigma\leq2$ controls the degree of nonlocality.
Throughout this work, we are mainly interested in the marginal case $\sigma=1$.

On a periodic domain, the fractional Laplacian is conveniently defined through its Fourier multiplier~\cite{DiNezzaPalatucciValdinoci2012,Kwasnicki2017}.
Writing
\begin{align}
    q(x)
    =
    \frac{1}{\sqrt L}
    \sum_k q_k e^{ikx},
    \qquad
    k=\frac{2\pi n}{L},
    \quad n\in\mathbb Z,
\end{align}
the quadratic part of the Hamiltonian is
\begin{align}
    H_{\mathrm{quad}}[q]
    =
    \sum_k |k|^\sigma |q_k|^2.
    \label{eq:H_quad}
\end{align}

Thus, the fractional operator is characterized by the nonanalytic symbol $|k|^\sigma$.
The case $\sigma=2$ reduces to the usual local gradient energy, while $0<\sigma<2$ corresponds to a genuinely nonlocal spatial coupling.

\begin{color}{black}
We note the relationship between nonlocality from nonlinearity in Equation~\eqref{eq:H_continuum}.
The fractional Laplacian is a linear operator acting on the complex field $q$, and the corresponding energy is quadratic in $q$.
The nonlinearity remains in the local quartic potential $g(|q|^2-1)^2$, whose contribution to the variational derivative is $2g(|q|^2-1)q$.
Thus, the present model does not transfer nonlinear effects from the potential term to the kinetic term.
Instead, it modifies the spatial range of the quadratic coupling while retaining a local nonlinear term that penalizes amplitude deviations from $|q|=1$.

The physical meaning of this choice becomes transparent in real space.
For $0<\sigma<2$, the continuum quadratic form for a sufficiently regular periodic field can be written as
\begin{align}
    H_{\rm quad}[q]
    =
    \frac{c_\sigma}{2}
    \int_0^L dx \int_0^L dx'\,
    \mathcal K_{\sigma,L}(x-x')
    |q(x)-q(x')|^2,
    \label{eq:fractional_kernel}
\end{align}
where $c_\sigma>0$ is the normalization constant associated with the Fourier multiplier $|k|^\sigma$, and
\begin{align*}
    \mathcal K_{\sigma,L}(r)
    =
    \sum_{p\in\mathbb Z}
    \frac{1}{|r+pL|^{1+\sigma}}
\end{align*}
is the periodized fractional kernel.
This is the periodic counterpart of the standard real-space representation of the fractional quadratic form~\cite{DiNezzaPalatucciValdinoci2012}.

In the fixed-amplitude limit, $q(x)=e^{i\theta(x)}$,
Equation~\eqref{eq:fractional_kernel} becomes
\begin{align}
    H_{\rm phase}[\theta]
    =
    c_\sigma
    \int_0^L dx \int_0^L dx'\,
    \mathcal K_{\sigma,L}(x-x')
    \left\{
        1-\cos[\theta(x)-\theta(x')]
    \right\}.
    \label{eq:long_range_phase_coupling}
\end{align}

The fractional quadratic term therefore directly represents long-range XY-type phase coupling.
This phase-only expression retains the compactness of the phase; it does not require a small-phase-difference expansion.
Related fractional-operator representations are also used in continuum descriptions of long-range XY models~\cite{GiachettiDefenuRuffoTrombettoni2021}.

This mechanism differs from introducing a nonlocal density--density interaction of the form
$\int dx\,dx'\,V(x-x')|q(x)|^2|q(x')|^2$.
Such a term depends on the density but has no explicit dependence on the phase.
When combined with an ordinary local gradient energy, its fixed-uniform-amplitude limit retains the local phase energy $\int dx\,(\partial_x\theta)^2$ rather than Equation~\eqref{eq:long_range_phase_coupling}.
Although density fluctuations can indirectly affect phase correlations, these two forms of nonlocality describe different physical mechanisms.

Our choice is motivated by the aim of studying how direct long-range phase coupling affects equilibrium correlations and twist responses, while retaining a simple local amplitude potential.
The exponent $\sigma$ controls the long-wavelength quadratic dispersion.
The case $\sigma=1$ is particularly relevant because the Gaussian phase fluctuations then grow logarithmically with separation.
This property motivates the numerical study, but does not by itself establish a finite-temperature BKT transition.

The formulation also provides a simple spectral implementation: the nonlocal linear operator acts through multiplication by $|k|^\sigma$, whereas the nonlinear force is evaluated locally
in physical space.
No density-dependent nonlocal convolution is required in the force evaluation.
This computational convenience facilitates the analysis of finite-size correlations and twist responses.

For a finite system, we consider the Gibbs measure
\begin{align}
    d\mu_{L,T}(q)
    =
    \frac{1}{Z_{L,T}}
    \exp\left[-\frac{H[q]}{T}\right]Dq,
    \label{eq:Gibbs_measure}
\end{align}
where $T$ is the temperature,
and $Z_{L,T}$ is the partition function.
All equilibrium averages discussed below are taken with respect to this measure.
\end{color}

\subsection{Spectral Twist and Fractional Twist Response}

In the two-dimensional XY model, the helicity modulus is defined through the free-energy cost of imposing a phase twist~\cite{FisherBarberJasnow1973,NelsonKosterlitz1977}.
For the present fractional model, the definition of a twist response requires some care because the spatial operator is nonlocal.
Here we use a spectral definition.

Let $\Delta$ be the total phase twist imposed across the system and define
\begin{align}
    \delta = \frac{\Delta}{L}.
\end{align}

For a local gradient energy, this corresponds to the replacement $\partial_x\to \partial_x+i\delta$.
For the fractional model, we define the twisted quadratic energy by shifting the Fourier multiplier,
\begin{align}
    H_{\mathrm{quad}}^{(\delta)}[q]
    =
    \sum_k |k+\delta|^\sigma |q_k|^2.
    \label{eq:H_quad_twisted}
\end{align}

Because the operator is nonlocal, this spectral twist should be regarded as one natural probe of phase rigidity rather than the unique definition of helicity modulus.
The full twisted Hamiltonian is then
\begin{align}
    H^{(\delta)}[q]
    =
    \sum_k |k+\delta|^\sigma |q_k|^2
    +
    g
    \int_0^L dx\,
    \left(|q(x)|^2-1\right)^2.
    \label{eq:H_twisted}
\end{align}
\begin{color}{black}

With this specified spectral prescription, the explicit twist dependence is entirely contained in the known Fourier multiplier.
The local potential is unchanged because it depends only on $|q|^2$.
This separation makes it possible to examine directly how the nonanalytic dispersion influences the twist free energy.
The interaction nevertheless affects the response through the equilibrium Gibbs measure.
We note that a nonlocal density--density interaction would also be invariant under a phase twist; therefore, the present advantage is the explicit representation of the long-range phase coupling, rather than a unique ability to define a twist.
\end{color}
The corresponding free energy is
\begin{align}
    F_L(\delta)
    =
    -T
    \log
    \int
    \exp\left[-\frac{H^{(\delta)}[q]}{T}\right]Dq,
\end{align}
and the twist free-energy difference is
\begin{align}
    \Delta F_L(\delta)
    =
    F_L(\delta)-F_L(0).
    \label{eq:DeltaF_def}
\end{align}

Equivalently, $\Delta F_L(\delta)$ can be evaluated by free-energy perturbation, or reweighting, with respect to the untwisted Gibbs measure~\cite{Zwanzig1954,FerrenbergSwendsen1988}:
\begin{align}
    \Delta F_L(\delta)
    =
    -T
    \log
    \left\langle
    \exp\left[
    -\frac{H^{(\delta)}[q]-H^{(0)}[q]}{T}
    \right]
    \right\rangle_{0},
    \label{eq:DeltaF_reweighting}
\end{align}
where $\langle \cdots\rangle_0$ denotes the expectation value for $H^{(0)}$.
In practice, this reweighting formula is reliable only when the ensembles for $H^{(0)}$ and $H^{(\delta)}$ have sufficient overlap.
Thermodynamic integration can also be used as a consistency check.

The nonanalyticity of the fractional dispersion immediately changes the small-twist behavior.
For a uniform configuration $q(x)=1$, only the zero Fourier component contributes.
Since $q_0=\sqrt L$, we obtain
\begin{align}
    H^{(\delta)}[1]-H^{(0)}[1]
    =
    L|\delta|^\sigma
    =
    L^{1-\sigma}|\Delta|^\sigma.
    \label{eq:uniform_twist_cost}
\end{align}

Thus, the natural twist response of the fractional model is not generally quadratic in $\delta$.
For $\sigma=2$, Equation~\eqref{eq:uniform_twist_cost} reduces to the usual analytic response proportional to $L\delta^2$.
For $\sigma=1$, however, the response is cusp-like and proportional to $|\Delta|$.

This motivates the finite-size quantity
\begin{align}
    Y_\sigma(L,T)
    =
    \frac{\Delta F_L(\delta)}
    {L|\delta|^\sigma}
    =
    \frac{\Delta F_L(\delta)}
    {L^{1-\sigma}|\Delta|^\sigma}.
    \label{eq:Ysigma_def}
\end{align}

For the marginal case $\sigma=1$, this becomes
\begin{align}
    Y_1(L,T)
    =
    \frac{\Delta F_L(\delta)}{|\Delta|}.
    \label{eq:cusp_response}
\end{align}

Although this quantity is a natural cusp-like analogue of the helicity modulus, it should not be interpreted a priori as the direct analogue of the helicity modulus in the two-dimensional XY model.
As will be examined numerically below, this quantity does not behave as a conventional helicity modulus.

\subsection{Spin-Wave Approximation and Absence of True Long-Range Order at $\sigma=1$}

To understand the role of long-wavelength fluctuations, it is useful to consider the spin-wave approximation.
In the low-temperature regime, amplitude fluctuations are suppressed and the complex field may be written as
\begin{align}
    q(x)\simeq e^{i\theta(x)}.
\end{align}

We expand the real phase field as
\begin{align}
    \theta(x)
    =
    \frac{1}{\sqrt L}
    \sum_k \theta_k e^{ikx},
    \qquad
    k=\frac{2\pi n}{L}.
\end{align}

Since $\theta(x)$ is real, $\theta_{-k}=\theta_k^\ast$.
The zero mode corresponds to a global $U(1)$ phase and does not contribute to phase differences.

Keeping only the quadratic phase fluctuation, we obtain the spin-wave Hamiltonian
\begin{align}
    H_{\mathrm{sw}}
    =
    \frac{\rho_{\rm s}}{2}
    \sum_{k\neq0}
    |k|^\sigma |\theta_k|^2,
    \label{eq:Hsw}
\end{align}
where $\rho_{\rm s}$ is an effective stiffness.
In the normalization of Equation~\eqref{eq:H_quad}, the bare spin-wave value of $\rho_{\rm s}$ is a matter of convention; the important point is the infrared dependence $|k|^\sigma$.

The Gaussian Gibbs measure for the spin-wave theory is
\begin{align}
    d\mu_{\rm sw}
    =
    \frac{1}{Z_{\rm sw}}
    \exp\left[-\frac{H_{\rm sw}}{T}\right]
    D\theta .
\end{align}

Equivalently, with $\beta=1/T$,
\begin{align}
    d\mu_{\rm sw}
    =
    \frac{1}{Z_{\rm sw}}
    \exp\left[-\beta H_{\rm sw}\right]
    D\theta .
\end{align}

For each independent Fourier mode, the Gaussian integral gives
\begin{align}
    \left\langle |\theta_k|^2\right\rangle_{\rm sw}
    \propto
    \frac{T}{\rho_{\rm s}|k|^\sigma}.
    \label{eq:theta_fluctuation}
\end{align}

The proportionality constant depends on the Fourier convention and on whether positive and negative wave numbers are counted independently.
The scaling $\langle |\theta_k|^2\rangle\propto |k|^{-\sigma}$ is independent of these conventions.

The equal-time correlation function is approximated by
\begin{align}
    C(r)
    =
    \left\langle
    e^{i[\theta(r)-\theta(0)]}
    \right\rangle
    \simeq
    \exp\left[
    -\frac12
    \left\langle
    (\theta(r)-\theta(0))^2
    \right\rangle
    \right],
    \label{eq:Gaussian_correlation}
\end{align}
where we used the standard identity for a Gaussian variable $X$,
\begin{align}
    \left\langle e^{iX}\right\rangle
    =
    \exp\left[-\frac12\left\langle X^2\right\rangle\right].
\end{align}

Using Equation~\eqref{eq:theta_fluctuation}, the phase-difference variance behaves as
\begin{align}
    \left\langle
    (\theta(r)-\theta(0))^2
    \right\rangle
    \propto
    \frac{T}{\rho_{\rm s}}
    \int_0^\Lambda dk\,
    \frac{1-\cos(kr)}{k^\sigma},
    \label{eq:phase_variance_integral}
\end{align}
where $\Lambda$ is an ultraviolet cutoff.

Equation~\eqref{eq:phase_variance_integral} shows the special role of $\sigma=1$.
For $\sigma<1$, the infrared fluctuation is finite and true long-range order is not excluded by the spin-wave approximation.
For $\sigma=1$, the integral grows logarithmically,
\begin{align}
    \left\langle
    (\theta(r)-\theta(0))^2
    \right\rangle
    \propto
    \frac{T}{\rho_{\rm s}}\log r,
\end{align}
and therefore
\begin{align}
    C(r)\sim r^{-\eta(T)}.
\end{align}

Thus, even in the low-temperature phase, the order parameter vanishes in the thermodynamic limit.
The low-temperature phase is not a phase with true spontaneous breaking of the continuous $U(1)$ symmetry, but rather a quasi-long-range ordered phase with algebraic correlations.
This behavior is analogous to the low-temperature phase of the two-dimensional XY model, but its origin is different: here the system is one-dimensional, and the logarithmic phase fluctuation arises from the marginal fractional dispersion $|k|$.
For $1 < \sigma \leq 2$, the same spin-wave approximation gives stronger-than-logarithmic phase fluctuations, leading to faster decay of correlations as
\begin{align}
    C(r) \sim \exp[-c r^{\sigma - 1}].
\end{align}

\subsection{Correlations in the Disordered Regime}

In short-range systems, the disordered phase is usually associated with exponential decay of correlations.
In the present nonlocal model, however, the situation is more subtle.
The nonanalyticity of the Fourier multiplier $|k|^\sigma$ can generate an algebraic tail even in the disordered regime.

To see this, consider an effective Gaussian Hamiltonian in the high-temperature or disordered phase,
\begin{align}
    H_{\rm G}
    =
    \sum_k
    \left(
    \mu^2+\rho_\sigma |k|^\sigma
    \right)
    |q_k|^2,
    \label{eq:HG_disordered}
\end{align}
where $\mu^2>0$ is an effective mass parameter and $\rho_\sigma>0$ is the coefficient of the fractional-gradient term in the Gaussian propagator.
\begin{color}{black}
For the auxiliary Gaussian Hamiltonian in Equation~\eqref{eq:HG_disordered}, the structure factor is
\begin{align}
    S_{\rm G}(k)
    =
    \left\langle |q_k|^2\right\rangle_{\rm G}
    =
    \frac{T}{\mu^2+\rho_\sigma |k|^\sigma},
    \label{eq:Sk_disordered}
\end{align}
where $\langle\cdots\rangle_{\rm G}$ denotes a Gaussian Gibbs average.
This expression is exact for the stated Gaussian Hamiltonian, but it is only an effective approximation to the structure factor of the original nonlinear model.

For $\sigma=1$, the long-distance tail can be obtained from Equation~\eqref{eq:Sk_disordered}.
For this analytical purpose, consider the infinite-line continuum Gaussian propagator at nonzero separation $r>0$ as
\begin{align}
    \begin{aligned}
        C_{\rm G}(r)
        &=
        \int_{-\infty}^{\infty}\frac{dk}{2\pi}\,
        \frac{T e^{ikr}}{\mu^2+\rho_1|k|} \\
        &=
        \frac{T}{\pi}\int_0^\infty dk\,
        \frac{\cos(kr)}{\mu^2+\rho_1 k} \\
        &=
        \frac{T\rho_1}{\pi}
        \int_0^\infty ds\,
        \frac{s e^{-\mu^2s}}{r^2+\rho_1^2s^2} \\
        &\sim
        \frac{T\rho_1}{\pi\mu^4}\frac{1}{r^2},
        \qquad r\to\infty.
    \end{aligned}
    \label{eq:rminus2_tail}
\end{align}

The scale separating the short- and long-distance behaviors of this
Gaussian propagator is $\xi_{\rm G}=\rho_1/\mu^2$;
the asymptotic expression applies for $r\gg\xi_{\rm G}$.
Thus, the full Gaussian propagator produces the $r^{-2}$ tail,
which is not an artifact of truncating a small-wave-number expansion.
The prefactor is a prediction of the auxiliary Gaussian description,
not an exact amplitude asserted for the interacting model.
The derivation above concerns an infinite-line continuum propagator;
quantitative comparisons with finite-grid results must also account
for the periodic geometry and the Fourier cutoff.

Neither Equation~\eqref{eq:Sk_disordered} nor a small-wave-number expansion
of it is used to generate the numerical data.
The reported correlations are measured directly from the equilibrium
configurations of the full nonlinear model, sampled with the stochastic
Gross--Pitaevskii dynamics at the specified finite cutoff.
The Gaussian argument is used only to interpret the possible
long-distance tail.
\end{color}

This algebraic tail should not be confused with quasi-long-range order.
In the disordered phase, the $r^{-2}$ decay is induced by the nonlocal fractional kernel and does not imply spontaneous ordering or vortex-pair binding in the same sense as in the BKT phase.
Close to the transition, the approach
 to this asymptotic tail may be governed by a very large crossover length.
Consequently, in finite-size simulations, the correlation function may exhibit intermediate-distance behavior that appears faster than a power law before ultimately crossing over to the nonlocal $r^{-2}$ tail at sufficiently large distances.

In contrast, below the transition temperature, the dominant long-distance correlation is expected to take the form
\begin{align}
    C(r)
    \sim
    r^{-\eta(T)},
    \qquad
    \eta(T)<2,
    \label{eq:lowT_algebraic}
\end{align}
which decays more slowly than the nonlocal high-temperature tail.
The transition can therefore be viewed as the emergence of a slower algebraic correlation on top of the background algebraic tail imposed by the fractional kernel.

\subsection{Observables and Finite-Size Scaling}

To characterize the transition, we use several dimensionless quantities.
The order parameter is defined by the zero Fourier component,
\begin{align}
    m
    =
    \left|
    \frac{1}{L}
    \int_0^L dx\,
    q(x)
    \right|.
    \label{eq:order_parameter}
\end{align}

Since true long-range order is absent at $\sigma=1$, $m$ is expected to vanish in the thermodynamic limit even in the low-temperature phase.
Nevertheless, its finite-size scaling provides useful information on algebraic order.
If
\begin{align}
    C(r)\sim r^{-\eta(T)},
\end{align}
then
\begin{align}
    \left\langle m^2\right\rangle
    \sim
    \begin{cases}
        L^{-\eta(T)} & 0 < \eta(T) < 1 \\
        (\log L) / L & \eta(T) = 1 \\
        L^{-1} & \eta(T) > 1
    \end{cases}.
    \label{eq:m2_scaling}
\end{align}

We also compute the Binder ratio~\cite{Binder1981}:
\begin{align}
    U_L(T)
    =
    \frac{\left\langle |m|^4\right\rangle}
    {\left\langle |m|^2\right\rangle^2},
    \label{eq:Binder}
\end{align}
and correlation ratios such as
\begin{align}
    R_C(L,T)
    =
    \frac{C(L/4)}{C(L/8)}.
    \label{eq:correlation_ratio}
\end{align}

In an algebraically ordered phase,
\begin{align}
    C(r)\sim r^{-\eta(T)}
\end{align}
implies
\begin{align}
    R_C(L,T)
    \to
    2^{-\eta(T)},
    \label{eq:RC_lowT}
\end{align}
which is independent of system size $L$.
In the high-temperature regime, if the asymptotic nonlocal tail
\begin{align}
    C(r)\sim r^{-2}
\end{align}
is reached, the same ratio approaches
\vspace{-6pt}\begin{align}
    R_C(L,T)
    \to
    2^{-2}
    =
    \frac14.
    \label{eq:RC_highT}
\end{align}

The finite-size crossover between these regimes provides a sensitive diagnostic of the transition.

It is also useful to define an effective exponent from the correlation ratio:
\begin{align}
    \eta_{\rm eff}(L,T)
    =
    -
    \frac{\log R_C(L,T)}{\log 2}.
    \label{eq:eta_eff}
\end{align}

In the algebraically ordered phase, $\eta_{\rm eff}(L,T)$ approaches $\eta(T)$.
In the high-temperature regime where the nonlocal $r^{-2}$ tail is observed, it approaches $2$.

Because the disordered phase of the fractional model can possess an algebraic long-distance tail, the term ``correlation length'' must be used with care.
In this work, we use $\xi_\ast(T)$ to denote a crossover length controlling the approach to the asymptotic high-temperature tail.
Near a BKT-type transition, this crossover length may become very large~\cite{Kosterlitz1974,WeberMinnhagen1988,HaradaKawashima1997}.
If the crossover length obeys the BKT-type essential singularity
\vspace{-6pt}\begin{align}
    \log \xi_\ast(T)
    =
    \frac{b}{\sqrt{T-T_{\rm BKT}}}
    +
    O(1)
    \qquad
    (T\downarrow T_{\rm BKT}),
    \label{eq:crossover_length_BKT}
\end{align}
then the condition $L/\xi_\ast(T)=\mathrm{const.}$ implies
\begin{align}
    (T-T_{\rm BKT})(\log L)^2
    =
    \mathrm{const.}
    \label{eq:BKT_FSS_variable}
\end{align}

Thus, a collapse of dimensionless quantities as functions of
\begin{align}
    (T-T_{\rm BKT})(\log L)^2
\end{align}
provides evidence for a BKT-type essential singularity.

We emphasize that Equation~\eqref{eq:BKT_FSS_variable} is used here as a finite-size scaling hypothesis rather than as an assumption built into the model.
Moreover, this scaling form is mainly applied on the disordered side and in the immediate vicinity of the transition.
In the algebraically ordered phase, dimensionless quantities such as $R_C(L,T)$ and $U_L(T)$ are expected to become approximately size-independent when plotted directly as functions of $T$, because the low-temperature phase is critical throughout.

Finally, we examine the twist-response quantity $Y_\sigma(L,T)$ defined in Equation~\eqref{eq:Ysigma_def}.
In the standard two-dimensional XY model, the helicity modulus remains finite in the low-temperature phase and exhibits a universal jump at the BKT transition.
In the present fractional one-dimensional system, however, the cusp-like response $Y_1(L,T)$ does not behave as an independent stiffness.
Instead, it closely follows the finite-size scaling of $\langle m^2\rangle$ and therefore vanishes in the thermodynamic limit in the algebraically ordered phase.
This distinction is essential for understanding how the present BKT-like transition differs from the conventional two-dimensional case.

\section{Stochastic Gross--Pitaevskii Dynamics and Gibbs Measure}\label{sec3}

\subsection{Variational Derivative of the Fractional Energy}

We use stochastic dynamics as a sampling method for equilibrium properties.
In this section, we introduce a stochastic Gross--Pitaevskii-type dynamics whose invariant measure is the Gibbs measure associated with the fractional Ginzburg--Landau energy~\cite{Stoof1999,GardinerDavis2003,BlakieBradleyDavisBallaghGardiner2008,CockburnProukakis2012,DeBouardDebusscheFukuizumi2018}.

Starting from the Hamiltonian in Equation~\eqref{eq:H_continuum}, its variational derivative with respect to $q^\ast$ is
\vspace{-8pt}\begin{align}
    \frac{\delta H}{\delta q^\ast}
    =
    (-\Delta)^{\sigma/2}q
    +
    2g\left(|q|^2-1\right)q.
    \label{eq:variational_derivative}
\end{align}

In Fourier space, the fractional operator is diagonal:
\begin{align}
    \left[(-\Delta)^{\sigma/2}q\right]_k
    =
    |k|^\sigma q_k.
\end{align}

For a mathematically well-defined finite-dimensional dynamics, we first introduce a Fourier cutoff.
Let $P_\Lambda$ be the projection onto the modes $|k|\le \Lambda$, and write
\begin{align}
    q_\Lambda(x)
    =
    \frac{1}{\sqrt L}
    \sum_{|k|\le \Lambda}
    q_k e^{ikx}.
\end{align}

The cutoff Hamiltonian is denoted by
\vspace{-6pt}\begin{align}
    H_\Lambda(q)
    =
    \sum_{|k|\le\Lambda}
    |k|^\sigma |q_k|^2
    +
    g
    \int_0^L dx\,
    \left(|q_\Lambda(x)|^2-1\right)^2.
    \label{eq:H_cutoff}
\end{align}

All stochastic equations below should be understood at this finite cutoff level.
The continuum notation is used for readability.

\subsection{Stochastic Gross--Pitaevskii Equation}

We consider the stochastic Gross--Pitaevskii-type equation
\begin{align}
    \partial_t q
    =
    - (\Gamma + i) \frac{\delta H}{\delta q^\ast}
    +
    \sqrt{\Gamma T} (\xi_1(x,t) + i \xi_2(x, t)),
    \qquad
    \Gamma>0,
    \label{eq:SGPE_form}
\end{align}
where $\Gamma$ is the damping parameter and $T$ is the temperature.
The noise fields $\xi_a$ satisfy
\vspace{-6pt}\begin{align}
    \left\langle
    \xi_a(x,t)\xi_b(x',t')
    \right\rangle
    =
    \delta_{ab}
    \delta(x-x')
    \delta(t-t'),
    \qquad
    a,b=1,2.
    \label{eq:noise_correlation}
\vspace{-6pt}\end{align}\begin{color}{black}

The deterministic drift consists of the dissipative contribution $-\Gamma\,\delta H/\delta q^\ast$ and the Hamiltonian contribution $-i\,\delta H/\delta q^\ast$.
The noise amplitude satisfies the fluctuation--dissipation relation associated with the dissipative contribution.
Here the stochastic equation is used to sample the equilibrium Gibbs measure of the prescribed Hamiltonian, rather than to model a driven nonequilibrium steady state.
\end{color}

Equation~\eqref{eq:SGPE_form} makes clear that the stochastic Gross--Pitaevskii dynamics consists of two parts.
The dissipative part decreases the Hamiltonian and, together with the noise, drives the system toward thermal equilibrium.
The Hamiltonian part conserves the Hamiltonian and rotates the field in phase space.
Therefore, the latter changes the sampling dynamics but not the equilibrium Gibbs measure.

\subsection{Real-Component Form}

To show the invariant measure explicitly, it is convenient to write the complex field as
\begin{align}
    q(x,t)=u(x,t)+iv(x,t),
\end{align}
and introduce the real two-component field
\vspace{-6pt}\begin{align}
    Q(x,t)=
    \begin{pmatrix}
        u(x,t)\\
        v(x,t)
    \end{pmatrix}.
\end{align}

At finite Fourier cutoff, $Q$ is a finite-dimensional vector.
The stochastic Gross--Pitaevskii dynamics can be written in the form
\begin{align}
    dQ
    =
    \frac{1}{2} \left[
        -\Gamma \nabla_Q H_\Lambda(Q)
        +
        J\nabla_Q H_\Lambda(Q)
    \right]dt
    +
    \sqrt{\Gamma T}\,dW,
    \label{eq:real_SDE}
\end{align}
where $W$ is a standard real Brownian motion in the finite-dimensional phase space and
\begin{align}
    J=
    \begin{pmatrix}
        0 & 1\\
        -1 & 0
    \end{pmatrix}
\end{align}
is the constant antisymmetric matrix representing the Hamiltonian flow.
The coefficient $\Gamma$ does not affect the equilibrium distribution as long as $\Gamma>0$ and the noise amplitude satisfies the fluctuation--dissipation relation.

\subsection{Invariant Gibbs Measure}

We now show that the finite-dimensional Gibbs measure
\begin{align}
    d\mu_{\Lambda,T}(Q)
    =
    \frac{1}{Z_{\Lambda,T}}
    \exp\left[-\frac{H_\Lambda(Q)}{T}\right]dQ
    \label{eq:finite_Gibbs_measure}
\end{align}
is invariant under Equation~\eqref{eq:real_SDE}.

Let $P(Q,t)$ be the probability density of $Q$.
The Fokker--Planck equation associated with Equation~\eqref{eq:real_SDE} is
\vspace{-6pt}\begin{align}
    \partial_t P
    =
    \frac{1}{2} \nabla_Q\cdot
    \left[
        \Gamma
        \left(
            \nabla_Q H_\Lambda\,P
            +
            T\nabla_Q P
        \right)
    \right]
    - \frac{1}{2}
    \nabla_Q\cdot
    \left[
        J\nabla_Q H_\Lambda\,P
    \right].
    \label{eq:Fokker_Planck}
\end{align}

We insert the Gibbs density
\begin{align}
    P_{\rm eq}(Q)
    =
    \frac{1}{Z_{\Lambda,T}}
    \exp\left[-\frac{H_\Lambda(Q)}{T}\right].
\end{align}

Since
\begin{align}
    \nabla_Q P_{\rm eq}
    =
    -\frac{1}{T}
    P_{\rm eq}
    \nabla_Q H_\Lambda,
\end{align}
the dissipative part vanishes:
\begin{align}
    \nabla_Q H_\Lambda\,P_{\rm eq}
    +
    T\nabla_Q P_{\rm eq}
    =
    0.
\end{align}

The Hamiltonian part also vanishes.
Indeed,
\vspace{-6pt}\begin{align}
    \nabla_Q\cdot
    \left(
        J\nabla_Q H_\Lambda\,P_{\rm eq}
    \right)
    =
    P_{\rm eq}
    \mathrm{Tr}
    \left(
        J\nabla_Q^2 H_\Lambda
    \right)
    +
    \left(
        J\nabla_Q H_\Lambda
    \right)\cdot
    \nabla_Q P_{\rm eq}.
\end{align}

The first term is zero because $J$ is antisymmetric while $\nabla_Q^2H_\Lambda$ is symmetric.
The second term is also zero because
\vspace{-6pt}\begin{align}
    \left(
        J\nabla_Q H_\Lambda
    \right)
    \cdot
    \nabla_Q H_\Lambda
    =
    0.
\end{align}

Therefore,
\vspace{-6pt}\begin{align}
    \partial_t P_{\rm eq}=0,
\end{align}
and the Gibbs measure in Equation~\eqref{eq:finite_Gibbs_measure} is invariant.

This derivation shows that the stochastic Gross--Pitaevskii dynamics samples the same equilibrium Gibbs measure as a purely dissipative Langevin dynamics.
The Hamiltonian Gross--Pitaevskii part affects dynamical properties and relaxation times, but it does not change equilibrium static observables.

\subsection{Equilibrium Interpretation}

In this work, the stochastic Gross--Pitaevskii equation is used as an equilibrium sampler.
The quantities studied below, such as the order parameter, Binder ratio, correlation ratio, effective exponent, and twist response, are static observables evaluated with respect to the Gibbs measure.
We do not use the stochastic dynamics to infer dynamical critical exponents or real-time transport properties.

This distinction is important because different stochastic dynamics can possess the same invariant Gibbs measure while having different dynamical behavior.
For example, underdamped Langevin dynamics, overdamped Langevin dynamics, and stochastic Gross--Pitaevskii dynamics may all sample the same equilibrium distribution if the fluctuation--dissipation relation is satisfied.
The results presented in this paper should therefore be interpreted as equilibrium properties of the fractional $U(1)$ Gibbs measure, rather than as properties specific to a particular stochastic time evolution.

\section{Numerical Results}\label{sec4}

\subsection{Numerical Method}

We numerically investigate the equilibrium properties of the one-dimensional fractional $U(1)$ model introduced in the previous section.
The system is defined on a periodic interval of length $L$, which is discretized as
\vspace{-6pt}\begin{align}
    x_j=j\Delta x,
    \qquad
    j=0,1,\ldots,N-1,
    \qquad
    L=N\Delta x .
\end{align}

In the simulations reported below, we set $\Delta x=1$, so that the system size is identified with the number of grid points $L=N$.

\textls[-25]{The field $q_j = q(x_j)$ is represented using a Fourier pseudospectral discretization~\cite{Trefethen2000,CanutoHussainiQuarteroniZang2006},}
\vspace{-6pt}\begin{align}
    q_j
    =
    \frac{1}{\sqrt L}
    \sum_{k\in \mathcal K_N}
    q_ke^{ikx_j},
\end{align}
where
\vspace{-6pt}\begin{align}
    \mathcal{K}_N = \left\{\frac{2\pi n}{L}: n = -\frac{N}{2}, \cdots, \frac{N}{2} - 1\right\},
\end{align}
and $\mathcal K_N$ denotes the set of Fourier modes retained by the finite discretization.
The cutoff Hamiltonian \eqref{eq:H_cutoff} is numerically evaluated by
\vspace{-6pt}\begin{align}
    H_\Lambda(q)
    =
    \sum_{k\in\mathcal{K}_N}
    |k|^\sigma |q_k|^2
    +
    g \Delta x \sum_{j = 0}^{N-1}
    \left(|q_j|^2-1\right)^2.
\end{align}

The equilibrium Gibbs measure is sampled by the stochastic Gross--Pitaevskii-type equation
\vspace{-6pt}\begin{align}
    \begin{aligned}
        \partial_t q_k
        &=
        -(i+\Gamma)
        \frac{\partial H_\Lambda(q)}{\partial q_k^\ast}
        +
        \sqrt{\Gamma T}\,
        \left(\xi_{1,k}+i\xi_{2,k}\right) \\
        &= -(i + \Gamma) \{ |k|^\sigma q_k + 2 g [(|q_j|^2 - 1) q_j]_k\} + \sqrt{\Gamma T}\, \left(\xi_{1,k} + i \xi_{2,k}\right),
        \qquad
        \Gamma>0.
    \end{aligned}
\end{align}

The nonlinear force on the right-hand side is calculated at each time step.
In the actual implementation, the Gaussian noises are generated in physical space as
\begin{align}
    \langle \xi_{a,j}(t)\xi_{b,j^\prime}(t')\rangle
    =
    \frac{1}{\Delta x}
    \delta_{ab}\delta_{jj^\prime}\delta(t-t').
\end{align}

They are then transformed to Fourier space using the same FFT convention as the field.
In this paper, we set $g=1$ and $\Gamma = 1$.
All quantities discussed below are equilibrium static observables.

\textls[-15]{The time integration is performed using a second-order stochastic Runge--Kutta scheme.
The time step is fixed to $\Delta t=0.01$.
For each temperature, the initial condition is chosen as the uniform ordered state $q_j=1$.
Starting from this configuration, the stochastic equation is integrated up to $t_{\rm eq}=1000$ for equilibration.
After equilibration, physical observables are measured every $\Delta t_{\rm sample}=1$.
For each temperature and system size, we collect $N_{\rm sample}=2^{23}$ measurements and compute ensemble averages from these samples.}

The principal observables are the equal-time correlation function, the Binder ratio, and the cusp twist response.
The order parameter is defined by the zero Fourier mode,
\vspace{-6pt}\begin{align}
    m
    =
    \left|
    \frac{1}{L}
    \int_0^L q(x)\,dx
    \right|
    \simeq
    \left|
    \frac{1}{N}
    \sum_{j=0}^{N-1} q_j
    \right|.
\end{align}

The Binder ratio is computed as
\vspace{-6pt}\begin{align}
    U_L(T)
    =
    \frac{\langle |m|^4\rangle}
    {\langle |m|^2\rangle^2}.
\end{align}

The two-point correlation function is evaluated as
\vspace{-6pt}\begin{align}
    C(r)
    =
    \frac{1}{L}
    \int_0^L dx\,
    \left\langle
    q^\ast(x)q(x+r)
    \right\rangle .
\end{align}

On the lattice, this becomes
\vspace{-6pt}\begin{align}
    C(r_\ell)
    =
    \frac{1}{N}
    \sum_{j=0}^{N-1}
    \left\langle
    \operatorname{Re}\left[q_j^\ast q_{j+\ell}\right]
    \right\rangle ,
    \qquad
    r_\ell=\ell\Delta x .
\end{align}

We mainly use the correlation ratio
\begin{align}
    R_C(L,T)
    =
    \frac{C(L/4)}{C(L/8)}
\end{align}
and the corresponding effective exponent
\vspace{-6pt}\begin{align}
    \eta_{\rm eff}(L,T)
    =
    -
    \frac{\log R_C(L,T)}{\log 2}.
\end{align}

If $C(r)\sim r^{-\eta(T)}$, then $\eta_{\rm eff}(L,T)$ approaches $\eta(T)$ in the large-$L$ limit.

The cusp twist response is evaluated from the spectral twist free-energy difference.
For a total twist angle $\Delta$, we set
\vspace{-6pt}\begin{align}
    \delta=\frac{\Delta}{L}
\end{align}
and define the twisted quadratic energy by
\begin{align}
    H_{\rm quad}^{(\delta)}
    =
    \sum_{k\in\mathcal{K}_N} |k+\delta|^\sigma |q_k|^2 .
\end{align}

The free-energy difference is computed by reweighting~\cite{Zwanzig1954,FerrenbergSwendsen1988}:
\begin{align}
    \Delta F_L(\delta)
    =
    -T
    \log
    \left\langle
    \exp\left[
    -\frac{H^{(\delta)}[q]-H^{(0)}[q]}{T}
    \right]
    \right\rangle_0.
\end{align}

This reweighting formula is used only for sufficiently small total twists for which the overlap between the two ensembles is adequate.
For $\sigma=1$, the corresponding cusp response is
\vspace{-6pt}\begin{align}
    Y_1(L,T)
    =
    \frac{\Delta F_L(\delta)}{|\Delta|}.
\end{align}

As discussed below, this quantity is useful for comparison with the helicity modulus of the two-dimensional XY model, but it does not remain finite in the thermodynamic limit in the algebraically ordered phase.

\subsection{Typical Stochastic Trajectories}

\begin{figure}[H] 
    \centering
    \includegraphics[width=0.8\linewidth]{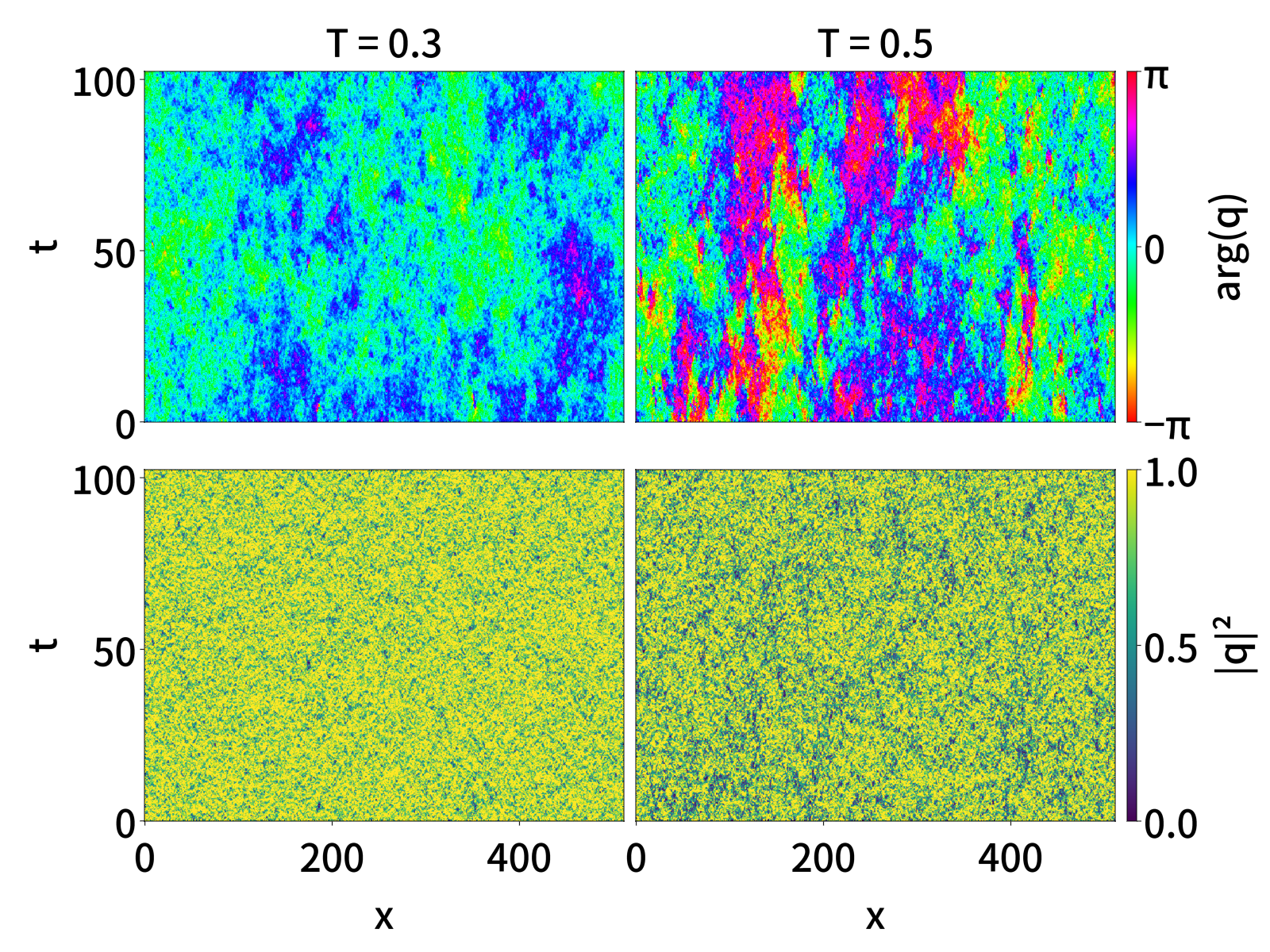}
    \caption{
    Typical 
 space-time plots of the phase and density obtained from the stochastic Gross--Pitaevskii dynamics.
    The upper panels show $\arg q(x,t)$ modulo $2\pi$, and the lower panels show $|q(x,t)|^2$.
    \textls[-15]{The left column corresponds to $T=0.3$, and the right column corresponds to $T=0.5$.
    The low-temperature trajectory shows slowly varying phase textures, whereas the high-temperature trajectory exhibits stronger phase fluctuations.
    This supports the interpretation that the dominant change across the transition is associated with phase fluctuations rather than large-scale amplitude ordering.}
    }
    \label{fig:snapshot}
\end{figure}

Before presenting the finite-size scaling analysis, we first show typical stochastic trajectories of the field $q(x,t)$.
Figure~\ref{fig:snapshot} shows space-time plots of the phase $\arg q(x,t)$ and the density $|q(x,t)|^2$.
The phase is plotted modulo $2\pi$, so apparent discontinuities can arise from the branch cut of the phase representation.
The horizontal axis is the spatial coordinate $x$, and the vertical axis is the simulation time $t$ after equilibration.
The left column corresponds to $T=0.3$, which is below the estimated transition temperature, while the right column corresponds to $T=0.5$, which is above the transition region.
The upper panels show $\arg q(x,t)$ modulo $2\pi$, and the lower panels show $|q(x,t)|^2$.

\textls[-15]{The most visible difference between the two temperatures appears in the phase field.
At $T=0.3$, the phase configuration exhibits slowly varying space-time textures.
Large correlated regions extend over long spatial and temporal scales, reflecting the presence of strong phase coherence in the low-temperature regime.
Although the phase is not uniformly ordered, its variation is smooth on large scales, which is consistent with quasi-long-range order.}

At $T=0.5$, the phase fluctuates much more strongly.
The phase plot contains rapidly varying structures, and the coherent space-time textures seen at $T=0.3$ become strongly fragmented.
This behavior is consistent with the loss of the low-temperature algebraic branch of the correlation function and with the strong size dependence observed in the correlation ratio near and above the transition region.

\textls[-15]{In contrast, the density $|q(x,t)|^2$ remains close to unity in both temperature regimes.
Although local density fluctuations are present, no large amplitude-domain structure appears across the transition.
This indicates that the dominant change across the transition is not an amplitude instability, but rather a change in the long-distance behavior of the phase degree of freedom.
The observation justifies, at least qualitatively, the spin-wave-based interpretation in which the low-temperature behavior is mainly governed by phase fluctuations.}

We emphasize that Figure~\ref{fig:snapshot} is intended as a qualitative visualization of typical stochastic trajectories.
Since the vertical direction represents simulation time, the time correlations in the figure depend on the chosen stochastic Gross--Pitaevskii dynamics and should not be interpreted as universal dynamical properties.
The transition itself is characterized below by equilibrium quantities, such as the correlation ratio, Binder ratio, effective exponent, and cusp twist response.

\begin{color}{black}

\subsection{Validation of Equilibrium Sampling}
\label{subsec:equilibrium_sampling_validation}

Before analyzing the equilibrium observables, we examine the numerical sampling using three complementary energy-based checks: the initial relaxation of the mean energy, sensitivity to the integration time step $\Delta t$, and thermodynamic consistency of energy averages and fluctuations.
These tests complement the invariant-measure argument for the continuous-time stochastic Gross--Pitaevskii equation.
Throughout this subsection, $E(t) = H_\Lambda[q(t)]$ denotes the total energy of the discretized field.

\subsubsection{Initial Energy Relaxation}
We first consider $L=256$ and initialize the field in the uniform
configuration\linebreak \mbox{ $q(x,0)=1$}.
The time-dependent mean energy $\overline{E}(t;T)$ is obtained by averaging over $2^{10}$ stochastic realizations.
We compare it with $E_{\rm ref}(T)$, the reference equilibrium energy estimated from the long-time simulations.
Figure~\ref{fig:validation_energy_relaxation}a shows this comparison at $T = 0.5$.
The mean energy approaches the reference value before the discarded equilibration time $t_{\rm eq}=1000$ and remains close to it over the subsequent observation interval, with no appreciable systematic drift on the scale of the figure.

\begin{figure}[H]
    \begin{minipage}{0.49\linewidth}
        \centering
        \includegraphics[width=0.99\linewidth]{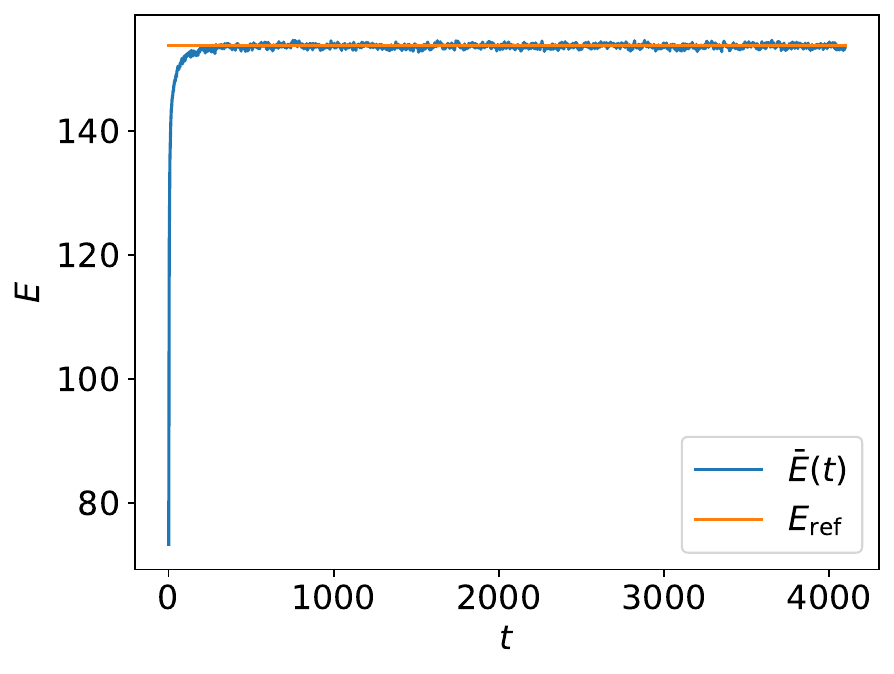} \\
        (\textbf{a})
    \end{minipage}
    \begin{minipage}{0.49\linewidth}
        \centering
        \includegraphics[width=0.99\linewidth]{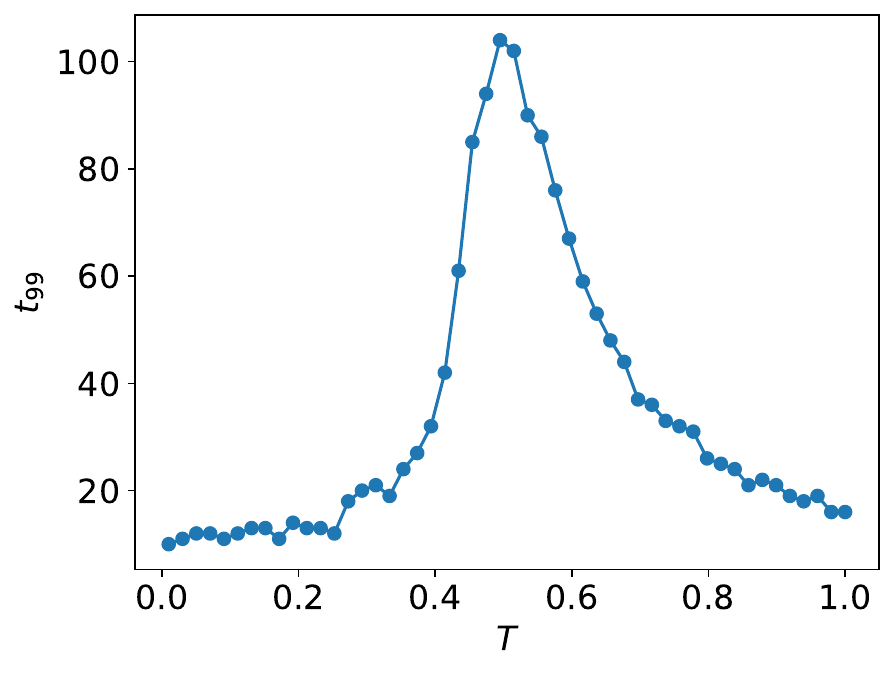} \\
        (\textbf{b})
    \end{minipage}
    \caption{\color{black}
    \label{fig:validation_energy_relaxation}
    Initial energy relaxation at $L=256$ from $q(x,0)=1$.
    (\textbf{a})  mean energy $\overline{E}(t)$ at $T=0.5$, averaged over $2^{10}$ stochastic realizations, compared with the reference equilibrium energy $E_{\rm ref}$ obtained from the long-time simulations.
    (\textbf{b})  temperature dependence of the first $99\%$ crossing time $t_{99}$ defined in Equation~\eqref{eq:validation_energy_t99}.
    The maximum is approximately $105$, compared with the discarded equilibration time $t_{\rm eq}=1000$.
    }
\end{figure}

To quantify the initial transient without assuming an exponential relaxation law, we define the threshold-crossing time
\begin{align}
    t_{99}(T)
    =
    \min\left\{
        t:\overline{E}(t;T)
        \geq 0.99 E_{\rm ref}(T)
    \right\}.
    \label{eq:validation_energy_t99}
\end{align}

This is the first crossing time of the mean energy, rather than the mean of the crossing times of individual trajectories.
Figure~\ref{fig:validation_energy_relaxation}b shows $t_{99}(T)$ over the temperature range examined.
Its maximum is approximately $105$, near $T=0.5$, and is therefore about an order of magnitude shorter than $t_{\rm eq}=1000$.
The discarded time is thus long compared with this operational measure of the initial energy transient at $L=256$.

\subsubsection{Time-Step Dependence}
We next examine the effect of halving the integration time step from $\Delta t=0.01$ to $\Delta t=0.005$ at $L=256$.
The comparison uses the fluctuation-based estimate of the heat capacity,
\begin{align}
    C_V^{\rm fluc}(T)
    =
    \frac{
        \langle E^2\rangle_T-\langle E\rangle_T^2
    }{T^2},
    \label{eq:validation_heat_capacity_fluc}
\end{align}
evaluated from the sampled energy moments for each time step.
Here $C_V$ denotes the heat capacity of the entire system.
The two curves in Figure~\ref{fig:validation_heat_capacity}a are nearly indistinguishable, including around the heat-capacity maximum.
The relative differences between their central estimates are at most approximately $1\%$ and are smaller than the estimated statistical uncertainties.
Thus, halving the time step produces no statistically resolved change in this energy-fluctuation observable at $L=256$.
This comparison tests time-step sensitivity at the available precision, rather than determining the remaining bias relative to the continuous-time limit.

\begin{figure}[H] 
    \begin{minipage}{0.49\linewidth}
        \centering
        \includegraphics[width=0.99\linewidth]{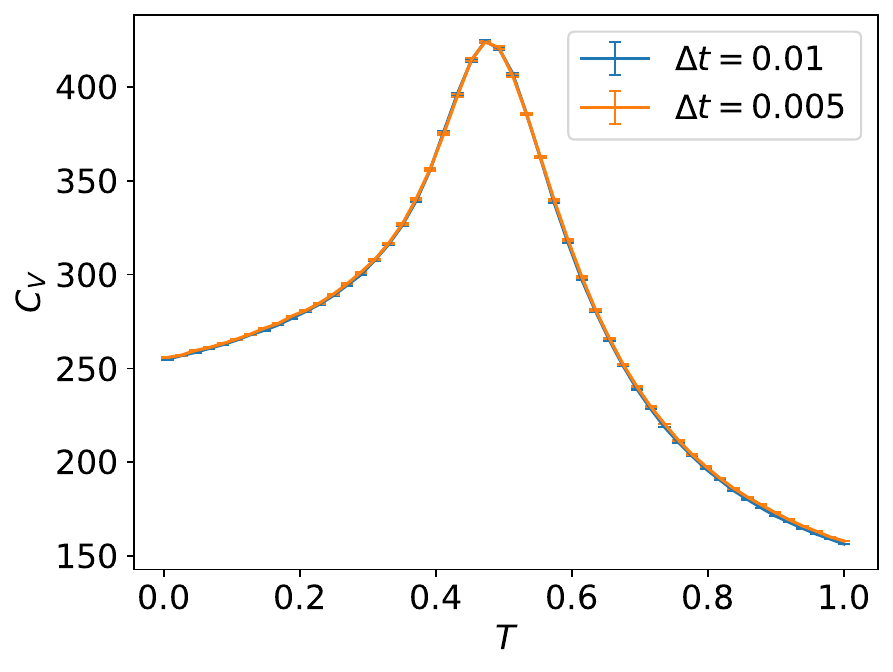} \\
        (\textbf{a})
    \end{minipage}
    \begin{minipage}{0.49\linewidth}
        \centering
        \includegraphics[width=0.99\linewidth]{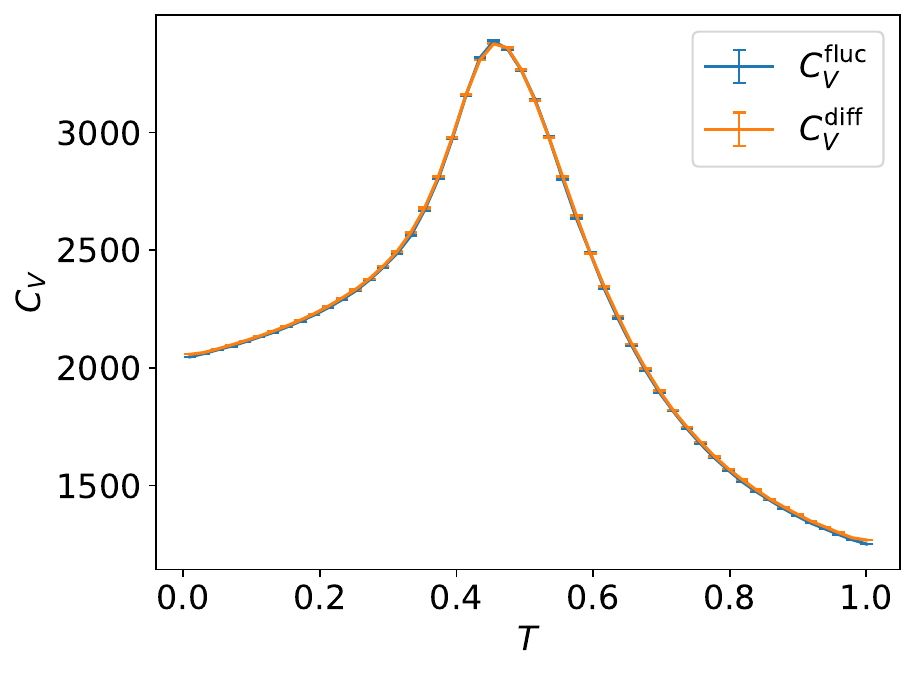} \\
        (\textbf{b})
    \end{minipage}
    \caption{\color{black}
    \label{fig:validation_heat_capacity}
    Heat-capacity checks of the numerical sampling.
    (\textbf{a})  Comparison of $C_V^{\rm fluc}$ obtained with $\Delta t=0.01$ and $\Delta t=0.005$ at $L=256$.
    (\textbf{b})  Comparison of the fluctuation-based estimate $C_V^{\rm fluc}$ and the central finite-difference estimate $C_V^{\rm diff}$ at $L=2048$.
    The definitions are given in \mbox{Equations~\eqref{eq:validation_heat_capacity_fluc} and \eqref{eq:validation_heat_capacity_diff}}.
    Error bars indicate estimated statistical standard errors.
    In both comparisons, the relative differences between the central estimates are approximately $1\%$ or less and are smaller than the estimated statistical uncertainties.
    }
\end{figure}

\paragraph{Thermodynamic Consistency}
As a complementary check, we use the available long-time data for $L=2048$, the largest system size studied.
For a Hamiltonian with no explicit temperature dependence, exact Gibbs averages at fixed $L$, Fourier cutoff, and coupling parameters satisfy
\begin{align}
    \frac{\partial\langle E\rangle_T}{\partial T} = \frac{\langle E^2\rangle_T-\langle E\rangle_T^2}{T^2}.
\end{align}

We therefore compare Equation~\eqref{eq:validation_heat_capacity_fluc} with the central finite-difference estimate
\begin{align}
    C_V^{\rm diff}(T)
    =
    \frac{
        \langle E\rangle_{T+h}-\langle E\rangle_{T-h}
    }{2h},
    \label{eq:validation_heat_capacity_diff}
\end{align}
where $h = 0.02$ is the temperature increment used in the centered stencil.
The latter estimate also involves a finite-temperature-step approximation to the derivative.

Figure~\ref{fig:validation_heat_capacity}a shows that $C_V^{\rm fluc}$ and $C_V^{\rm diff}$ agree within their estimated statistical uncertainties over the temperature range examined, including near the maximum.
Their relative differences are approximately $1\%$ or less and are smaller than the estimated statistical uncertainties.
This agreement supports the thermodynamic consistency of the sampled energy statistics at $L=2048$.

Together, these results provide complementary, energy-based checks of the equilibrium sampling procedure.
They assess the initial energy transient and time-step sensitivity at $L=256$, and thermodynamic consistency at $L=2048$.
\end{color}

\subsection{Correlation Function, Correlation Ratio, and Binder Ratio}

\begin{figure}[H] 
    \centering
    \includegraphics[width=0.5\linewidth]{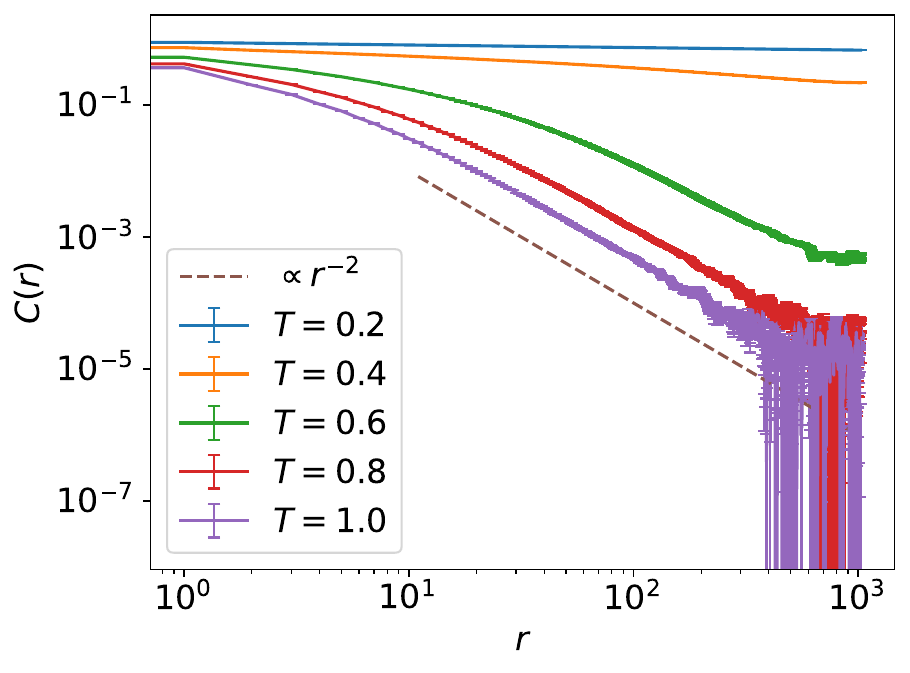}
    \caption{
    Correlation function $C(r)$ with $L=2048$ for several temperatures.
    At low temperatures, $C(r)$ decays slowly and is consistent with an algebraic form.
    At high temperatures, $C(r)$ decays rapidly at intermediate distances and becomes compatible with the fractional-kernel-induced algebraic tail $C(r)\sim r^{-2}$ at long distances.
    }
    \label{fig:correlation_function}
\end{figure}

We first examine the equilibrium correlation function defined above.
Figure~\ref{fig:correlation_function} shows $C(r)$ for several temperatures in a log--log plot.
At low temperatures, the decay is slow and is consistent with an algebraic form $C(r)\sim r^{-\eta(T)}$, with a temperature-dependent exponent that remains small in the low-temperature regime.
This behavior is consistent with quasi-long-range order.

\textls[-15]{At higher temperatures, the correlation function decays much more rapidly over intermediate distances.
However, the decay is not simply the ordinary exponential decay expected in short-range one-dimensional systems.
Instead, the data are consistent with the presence of a long-distance algebraic tail induced by the nonlocal fractional kernel.
For $\sigma=1$, the Gaussian argument in Section~\ref{sec:model} predicts the asymptotic high-temperature tail $C(r)\sim r^{-2}$, shown in Figure~\ref{fig:correlation_function} as a reference line.
The numerical results suggest that the approach to this $r^{-2}$ regime is controlled by a large crossover length, especially near the transition.}

\begin{figure}[H] 
    \begin{minipage}{0.49\linewidth}
        \centering
        \includegraphics[width=0.99\linewidth]{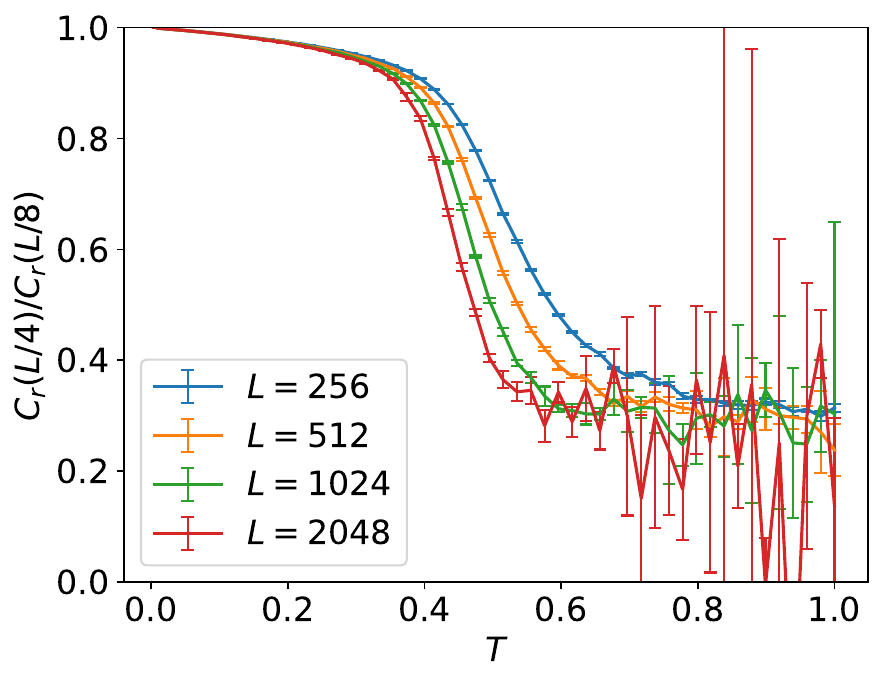} \\
        (\textbf{a})
    \end{minipage}
    \begin{minipage}{0.49\linewidth}
        \centering
        \includegraphics[width=0.99\linewidth]{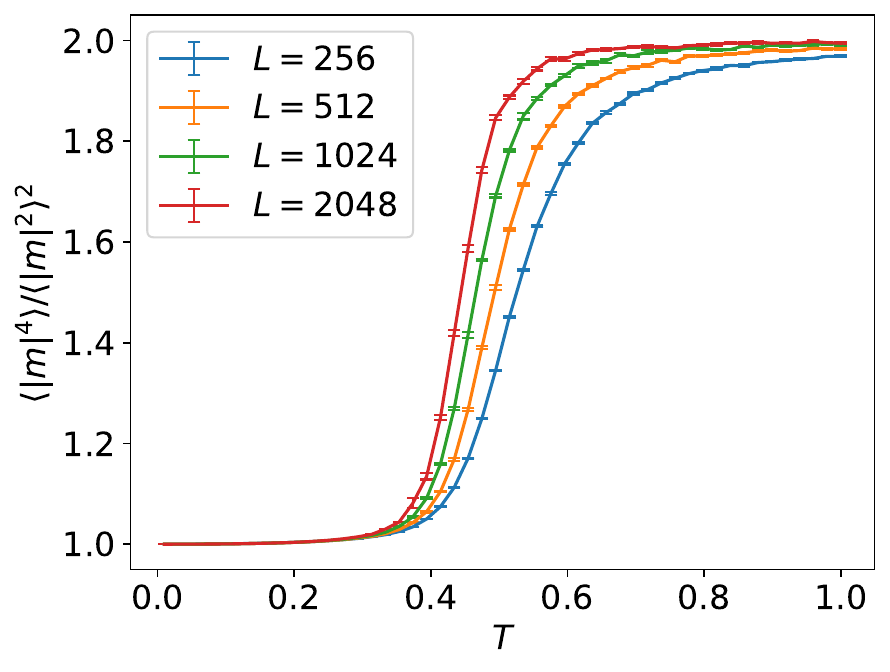} \\
        (\textbf{b})
    \end{minipage}
    \caption{
    Temperature dependence of (\textbf{a}) the correlation ratio $R_C(L,T)=C(L/4)/C(L/8)$ and (\textbf{b})~the Binder ratio $U_L(T)=\langle |m|^4\rangle/\langle |m|^2\rangle^2$ for $L=256,512,1024,2048$.
    Both quantities show weak size dependence in the low-temperature algebraic regime.
    The size dependence becomes pronounced near the transition region.
    }
    \label{fig:ratios}
\end{figure}

To quantify the change in the long-distance behavior, we use the correlation ratio $R_C(L,T)$ defined in Equation~\eqref{eq:correlation_ratio}.
In an algebraically ordered phase, $R_C(L,T)$ approaches $2^{-\eta(T)}$ and becomes approximately independent of system size.
Figure~\ref{fig:ratios} shows that the curves for different $L$ almost overlap at low temperatures, indicating a size-independent slow algebraic branch.
As the temperature increases, the curves start to separate, signaling the loss of this low-temperature branch.
At sufficiently high temperatures, the values of $R_C$ approach the vicinity of $1/4$, as expected from the nonlocal $r^{-2}$ tail.
Thus, the temperature dependence of $R_C$ suggests three regimes: a low-temperature algebraic regime, a crossover regime with strong finite-size dependence, and a high-temperature regime controlled by the nonlocal tail.

We also compute the Binder ratio $U_L(T)$ defined in Equation~\eqref{eq:Binder}.
The Binder ratio is close to unity in the low-temperature regime and is nearly independent of $L$, consistent with algebraic rather than truly disordered behavior.
Around the crossover region, $U_L(T)$ develops a strong size dependence, with an onset in approximately the same temperature range as that observed in the correlation ratio.
At high temperatures, $U_L(T)$ approaches the value $2$, which is expected when the complex zero mode behaves approximately as a Gaussian variable with zero mean.

The low-temperature size-independent branches of $R_C$ and $U_L$ begin to break down in the range $T\simeq0.35\text{--}0.4$.
We therefore use $T_{\rm BKT}\simeq0.35$ as a preliminary operational estimate, which will be refined by the finite-size scaling analysis below.
Because BKT-type finite-size corrections are logarithmically slow, a precise determination of $T_{\rm BKT}$ requires a finite-size scaling analysis.

\subsection{Twist Response and Helicity-Modulus-like Quantities}

We next examine twist-response quantities.
In the conventional two-dimensional XY model, the helicity modulus is one of the central observables for identifying the BKT transition.
It remains finite in the low-temperature phase and exhibits a universal jump at the transition.
It is therefore natural to ask whether an analogous stiffness can be defined in the present one-dimensional fractional model.

We first consider the most direct analogue of the ordinary helicity modulus.
For this purpose, we use the real-space nonlocal representation of the fractional quadratic energy and introduce a Peierls-type twist on each nonlocal link.
Specifically, we write
\begin{align}
H_{\rm nl}^{(\delta)}[q]
=
\sum_j
\sum_{r}
\alpha_r
\left|
q_j-q_{j+r}e^{ir\delta}
\right|^2
+
V[q],
\label{eq:H_nonlocal_twisted}
\end{align}
where $V[q]$ denotes the local nonlinear term.
The coefficients $\alpha_r$ are chosen so that the untwisted quadratic form reproduces the fractional dispersion,
\begin{align}
2
\sum_r
\alpha_r
\left(
1-\cos pr
\right)
\simeq
|p|^\sigma
\end{align}
for the Fourier modes of the finite system.
For $\sigma=1$, the corresponding real-space kernel is long-ranged and behaves asymptotically as $\alpha_r\sim r^{-2}$.

Expanding Equation~\eqref{eq:H_nonlocal_twisted} for small $\delta$, we obtain
\begin{align}
H_{\rm nl}^{(\delta)}
=
H_{\rm nl}^{(0)}
+
\delta A_{\rm nl}
+
\delta^2 B_{\rm nl}
+
O(\delta^3),
\end{align}
where
\begin{align}
A_{\rm nl}
=
2
\sum_j
\sum_r
r\alpha_r
\operatorname{Im}
\left(
q_j^\ast q_{j+r}
\right),
\label{eq:A_nonlocal}
\end{align}
and
\begin{align}
B_{\rm nl}
=
\sum_j
\sum_r
r^2\alpha_r
\operatorname{Re}
\left(
q_j^\ast q_{j+r}
\right).
\label{eq:B_nonlocal}
\end{align}

The corresponding linear-response helicity-modulus-like quantity is
\begin{align}
Y_{\rm LR}
=
\frac{1}{L}
\left[
\left\langle B_{\rm nl}\right\rangle
-
\frac{1}{2T}
\left(
\left\langle A_{\rm nl}^2\right\rangle
-
\left\langle A_{\rm nl}\right\rangle^2
\right)
\right].
\label{eq:Helicity_LR}
\end{align}

This definition is the direct analogue of the conventional helicity modulus formula, but it assumes that the twist free energy is analytic and quadratic in $\delta$.

\begin{figure}[H] 
\centering
\includegraphics[width=0.5\linewidth]{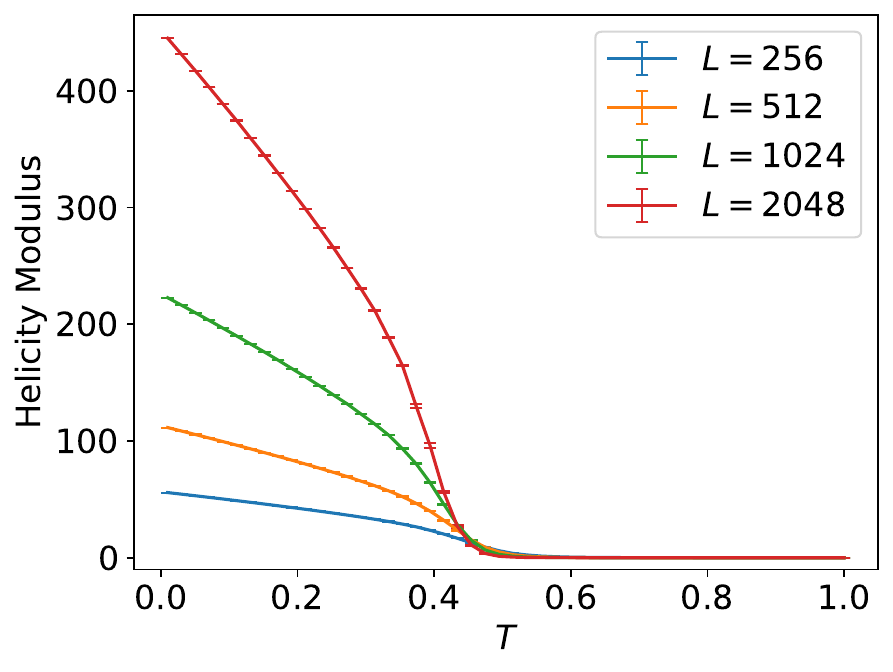}
\caption{
Helicity-modulus-like quantity obtained from the ordinary linear-response formula in Equation~\eqref{eq:Helicity_LR}.
This quantity is defined by applying a Peierls-type twist to the real-space nonlocal representation of the fractional quadratic energy, thereby assuming an analytic $\delta^2$ response.
In the low-temperature regime, the response grows strongly with system size and does not approach a finite size-independent value.
This shows that the ordinary helicity modulus is not an appropriate finite stiffness for the $\sigma=1$ fractional model.
}
\label{fig:linear_helicity}
\end{figure}

The result is shown in Figure~\ref{fig:linear_helicity}.
In the low-temperature regime, $Y_{\rm LR}$ grows strongly with system size.
The curves for $L=256,512,1024,2048$ do not approach a common finite value.
Instead, the magnitude of the response increases as $L$ is increased.
This behavior indicates that the usual analytic helicity modulus is not a well-defined finite stiffness for the present $\sigma=1$ fractional model.
The origin of this behavior is already visible at the level of the kernel.
For $\sigma=1$, $\alpha_r\sim r^{-2}$, and therefore the factor $r^2\alpha_r$ appearing in $B_{\rm nl}$ does not decay at large $r$.
Consequently, the ordinary quadratic twist response receives contributions from increasingly long nonlocal links as the system size is increased.

This failure of the ordinary linear-response stiffness is consistent with the nonanalytic form of the fractional dispersion.
For the spectral fractional model, the natural twisted quadratic energy is obtained by shifting the Fourier multiplier,
\begin{align}
H_{\rm quad}^{(\delta)}
=
\sum_{k\in\mathcal K_N}
|k+\delta|^\sigma |q_k|^2 .
\label{eq:spectral_twist_quad}
\end{align}

For a uniform configuration $q(x)=1$, only the zero Fourier mode contributes.
Since $q_0=\sqrt L$, one obtains
\begin{align}
H_{\rm quad}^{(\delta)}[1]
-
H_{\rm quad}^{(0)}[1]
=
L|\delta|^\sigma .
\end{align}

Thus, for $\sigma=1$, the natural zero-mode twist cost is proportional to
\begin{align}
L|\delta|
=
|\Delta|,
\end{align}
where $\Delta=L\delta$ is the total twist angle.
The response is therefore cusp-like rather than quadratic.

We therefore consider the cusp twist response defined from the spectral twist free-energy difference.
For a total twist angle $\Delta$, we set
\begin{align}
\delta=\frac{\Delta}{L},
\end{align}
and define
\begin{align}
\Delta F_L(\delta)
=
F_L(\delta)-F_L(0).
\end{align}

For $\sigma=1$, the natural finite-size response is
\begin{align}
Y_{\rm cusp}(L,T)
=
\frac{\Delta F_L(\delta)}{|\Delta|}.
\label{eq:Y_cusp}
\end{align}

In the data shown below, unless otherwise stated, we use $\Delta=\pi/4$.
The dependence on $\Delta$ will be discussed separately.

\begin{figure}[H] 
\begin{minipage}{0.49\linewidth}
    \centering
    \includegraphics[width=0.99\linewidth]{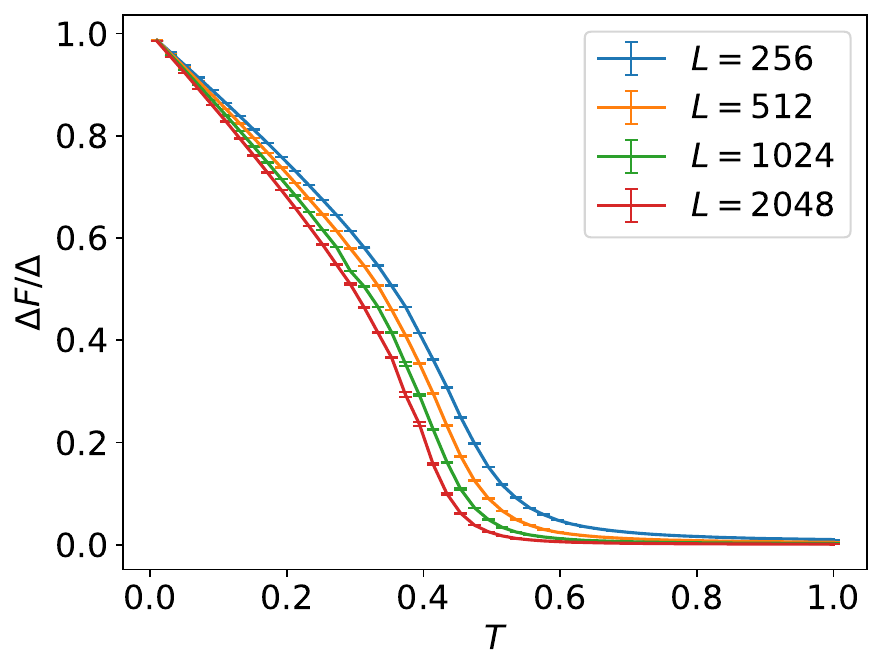} \\
    (\textbf{a})
\end{minipage}
\begin{minipage}{0.49\linewidth}
    \centering
    \includegraphics[width=0.99\linewidth]{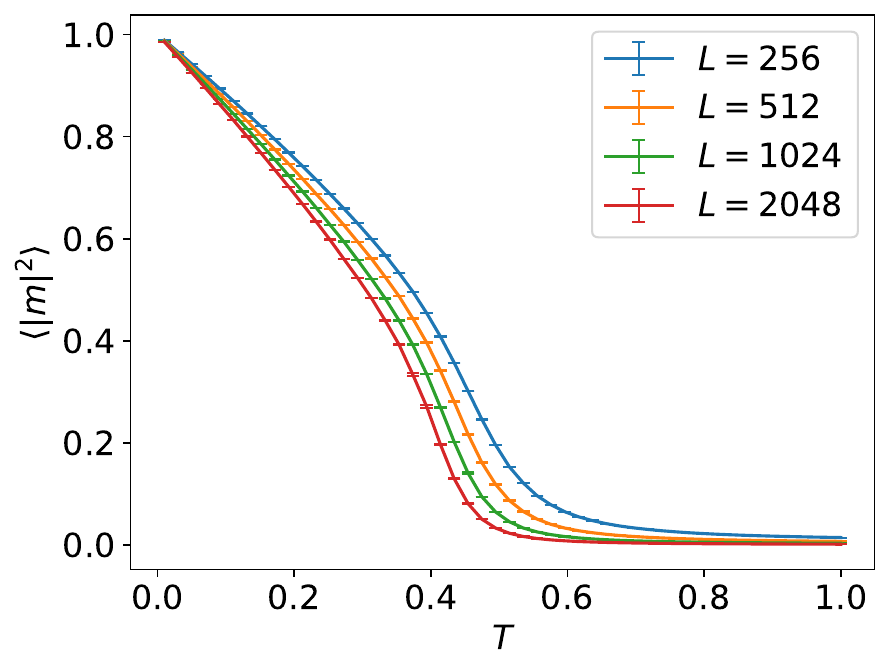} \\
    (\textbf{b})
\end{minipage}
\caption{
(\textbf{a}) Cusp twist response $Y_{\rm cusp}(L,T)=\Delta F_L/|\Delta|$ computed with $\Delta=\pi/4$, and (\textbf{b}) the squared order parameter $\langle |m|^2\rangle$.
Although $Y_{\rm cusp}$ remains finite at each finite system size, it decreases with increasing $L$ in the low-temperature regime.
Its temperature and size dependence closely resemble those of $\langle |m|^2\rangle$.
Therefore, $Y_{\rm cusp}$ does not behave as an independent stiffness analogous to the helicity modulus of the two-dimensional XY model.
}
\label{fig:cusp_helicity}
\end{figure}

Figure~\ref{fig:cusp_helicity} shows $Y_{\rm cusp}(L,T)$ and $\langle |m|^2\rangle$ as functions of temperature.
Unlike the ordinary linear-response helicity-modulus-like quantity, $Y_{\rm cusp}$ does not diverge with system size.
At each finite $L$, it decreases from a value close to unity at low temperature to nearly zero in the high-temperature regime.
However, in the low-temperature algebraic regime, the value of $Y_{\rm cusp}$ systematically decreases as $L$ is increased.
Thus, $Y_{\rm cusp}$ does not behave as a finite stiffness that survives in the thermodynamic limit.

The similarity between $Y_{\rm cusp}$ and $\langle |m|^2\rangle$ can be understood from the zero-mode contribution to the spectral twist.
For $\sigma=1$,
\begin{align}
H_{\rm quad}^{(\delta)}
-
H_{\rm quad}^{(0)}
=
\sum_{k\in\mathcal K_N}
\left(
|k+\delta|-|k|
\right)
|q_k|^2 .
\end{align}

The zero-mode contribution is
\begin{align}
\Delta H_0
=
|\delta| |q_0|^2 .
\end{align}

Using
\begin{align}
m
=
\left|
\frac{1}{L}
\int_0^L q(x) dx
\right|
=
\frac{|q_0|}{\sqrt L},
\end{align}
we obtain
\begin{align}
\Delta H_0
=
|\delta|L|m|^2
=
|\Delta| |m|^2 .
\label{eq:zeromode_twist}
\end{align}

Therefore, the cusp twist response contains a direct contribution from the squared zero-mode order parameter.
This provides a natural explanation for why $Y_{\rm cusp}(L,T)$ closely follows the finite-size behavior of $\langle |m|^2\rangle$.

In the low-temperature algebraic regime, the correlation function behaves as
\begin{align}
C(r)\sim r^{-\eta(T)}.
\end{align}

For the slow algebraic branch with $0<\eta(T)<1$, this implies
\begin{align}
\langle |m|^2\rangle
\sim
L^{-\eta(T)} .
\end{align}

The data in Figure~\ref{fig:cusp_helicity} suggest that $Y_{\rm cusp}(L,T)$ is governed by the same leading finite-size scaling.
Thus, although $Y_{\rm cusp}$ is a natural finite-twist response for the fractional spectral model, it is not an independent thermodynamic stiffness in the same sense as the helicity modulus of the conventional two-dimensional XY model.

These results show that neither the ordinary analytic linear-response quantity nor the spectral cusp response provides a direct analogue of the two-dimensional helicity modulus.
The former grows with system size because the nonlocal kernel gives long-distance contributions to the quadratic twist response.
The latter is strongly tied to the zero-mode order parameter and decreases together with $\langle |m|^2\rangle$ in the algebraically ordered regime.
Therefore, the BKT-like transition in the present fractional one-dimensional model must be characterized primarily through correlation functions and dimensionless finite-size scaling quantities, rather than through a universal jump of a conventional stiffness.

\subsection{Finite-Size Scaling of the Correlation and Binder Ratios}

We now examine the finite-size scaling of the dimensionless quantities introduced above.
Since the twist-response quantities do not provide an analogue of the two-dimensional helicity modulus, we characterize the transition mainly through the correlation ratio and the Binder ratio.
As discussed in Section~\ref{sec:model}, the BKT essential singularity~\cite{Kosterlitz1974,WeberMinnhagen1988,HaradaKawashima1997} suggests the scaling variable
\begin{align}
    X=(T-T_{\rm BKT})(\log L)^2 .
    \label{eq:BKT_scaling_variable_results}
\end{align}

The low-temperature size-independent branch starts to break down around $T\simeq0.35\text{--}0.4$.
Guided by this observation, we use $T_{\rm BKT}=0.35$ in the finite-size scaling analysis.
This value is not obtained by optimizing the collapse itself, but from the onset of finite-size dependence in the correlation and Binder ratios.
The collapse analysis therefore provides a nontrivial consistency check of the BKT-like interpretation.

\begin{figure}[H] 
    \begin{minipage}{0.49\linewidth}
        \centering
        \includegraphics[width=0.99\linewidth]{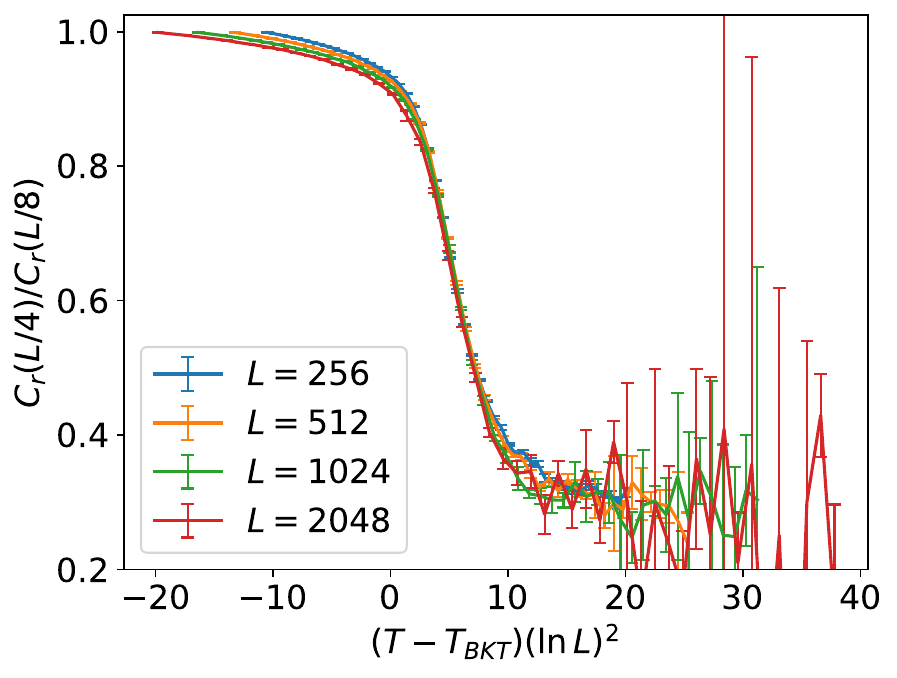} \\
        (\textbf{a})
    \end{minipage}
    \begin{minipage}{0.49\linewidth}
        \centering
        \includegraphics[width=0.99\linewidth]{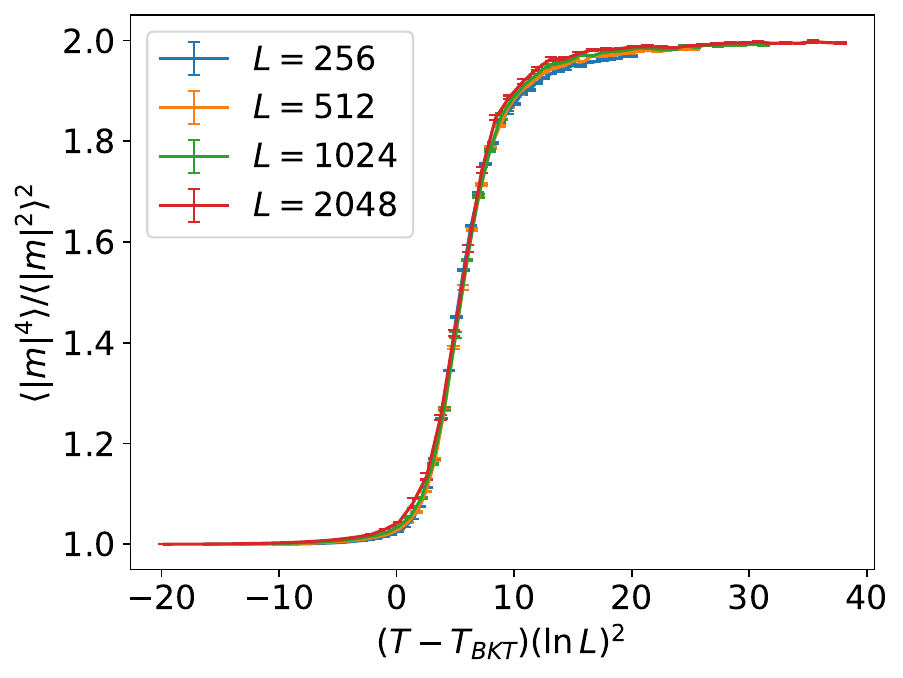} \\
        (\textbf{b})
    \end{minipage}
    \caption{
    Finite-size scaling of (\textbf{a}) the correlation ratio $R_C(L,T)=C(L/4)/C(L/8)$ and (\textbf{b}) the Binder ratio $U_L(T)=\langle |m|^4\rangle/\langle |m|^2\rangle^2$.
    The horizontal axis is $X=(T-T_{\rm BKT})(\log L)^2$, with $T_{\rm BKT}=0.35$.
    The data for $L=256,512,1024,2048$ collapse well in the crossover region, which is consistent with a BKT-type essential scaling of the crossover length.
    }
    \label{fig:Rc_scaling}
\end{figure}

Figures~\ref{fig:Rc_scaling}a,b show the scaling plots of $R_C$ and $U_L$, respectively.
The data for $L=256,512,1024,2048$ show a good collapse as functions of $X$, especially in the rapid crossover region where the curves separate when plotted directly as functions of $T$.
This supports the interpretation that the characteristic crossover length obeys a BKT-type essential singularity on the high-temperature side.
Possible additive logarithmic corrections, such as replacing $\log L$ by $\log L+c$, are not included in this minimal collapse analysis.

It is important to emphasize the meaning of this scaling collapse.
The low-temperature phase is algebraically ordered over an extended temperature range, so dimensionless ratios such as $R_C$ and $U_L$ are expected to become approximately size independent when plotted directly as functions of $T$.
The collapse with Equation~\eqref{eq:BKT_scaling_variable_results} should therefore be interpreted mainly as a scaling analysis of the disordered side and the crossover region near $T_{\rm BKT}$, rather than as an ordinary critical finite-size scaling over the entire temperature range.
Moreover, the length scale involved here should be understood as the crossover length $\xi_\ast(T)$ associated with the approach to the high-temperature asymptotic tail, not as a conventional exponential correlation length.

These results provide evidence that the observed transition is BKT-like in the sense of logarithmic finite-size scaling and the emergence of a low-temperature algebraic branch.
At the same time, as discussed in the previous subsection, the absence of an independent helicity-modulus-like stiffness distinguishes the present transition from the conventional two-dimensional XY transition.

\subsection{Effective Exponent and Scaling of the Cusp Response}

We next examine the effective exponent $\eta_{\rm eff}(L,T)$ defined in Equation~\eqref{eq:eta_eff}.
If the correlation function behaves algebraically, then $\eta_{\rm eff}(L,T)$ approaches the corresponding algebraic exponent in the large-$L$ limit.

\begin{figure}[H] 
    \centering
    \includegraphics[width=0.50\linewidth]{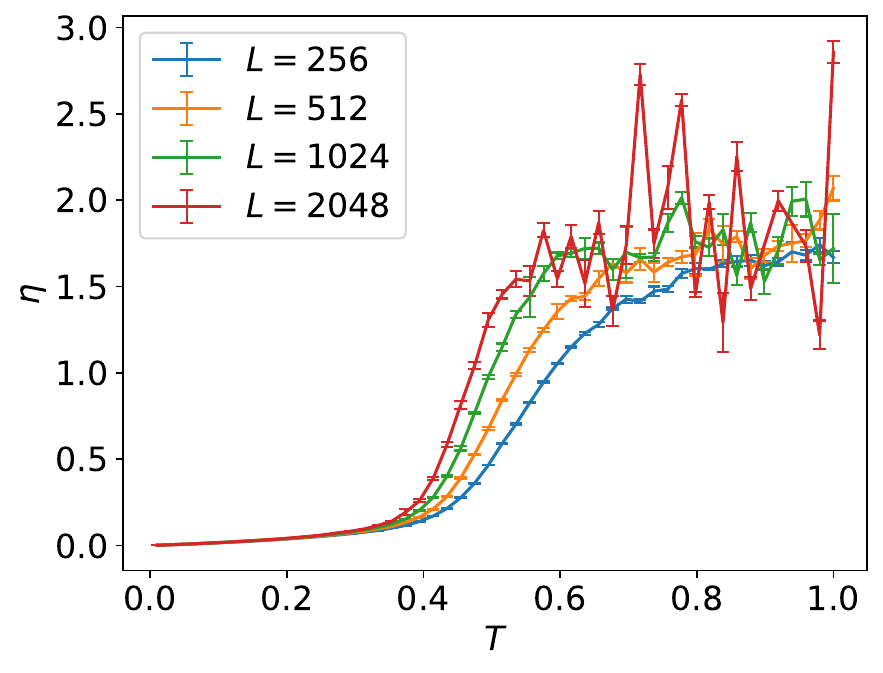}
    \caption{
    Effective exponent $\eta_{\rm eff}(L,T)=-\log[C(L/4)/C(L/8)]/\log 2$.
    In the low-temperature regime, $\eta_{\rm eff}$ is nearly independent of system size, indicating algebraic correlations.
    In the high-temperature regime, $\eta_{\rm eff}$ approaches $2$, consistent with the fractional-kernel-induced tail $C(r)\sim r^{-2}$.
    }
    \label{fig:eta_eff}
\end{figure}

Figure~\ref{fig:eta_eff} shows the temperature dependence of $\eta_{\rm eff}(L,T)$.
In the low-temperature regime, the curves for different system sizes almost overlap.
This indicates that the low-temperature phase is characterized by a size-independent algebraic exponent $\eta(T)$.
At higher temperatures, $\eta_{\rm eff}$ becomes strongly size-dependent in the crossover region and eventually approaches a value close to $2$.
This high-temperature value should not be interpreted as the critical exponent of the BKT-like transition; it corresponds to the asymptotic $r^{-2}$ tail induced by the nonlocal fractional kernel for $\sigma=1$.
Thus, the slow low-temperature algebraic branch and the high-temperature fractional-kernel-induced tail have different origins and should be distinguished.

The effective exponent also allows us to test the finite-size scaling of quantities related to the zero Fourier mode.
For the slow algebraic branch with $0<\eta(T)<1$, Equation~\eqref{eq:m2_scaling} implies $\langle |m|^2\rangle\sim L^{-\eta(T)}$.
Therefore, $\langle |m|^2\rangle L^{\eta(T)}$ should become independent of $L$ in the low-temperature algebraic regime.
In this scaling test, we use the finite-size estimate $\eta_{\rm eff}(L,T)$ obtained from the same correlation ratio.
The purpose is therefore to test the internal consistency between the correlation-ratio exponent and the zero-mode scaling.

\begin{figure}[H] 
    \begin{minipage}{0.49\linewidth}
        \centering
        \includegraphics[width=0.99\linewidth]{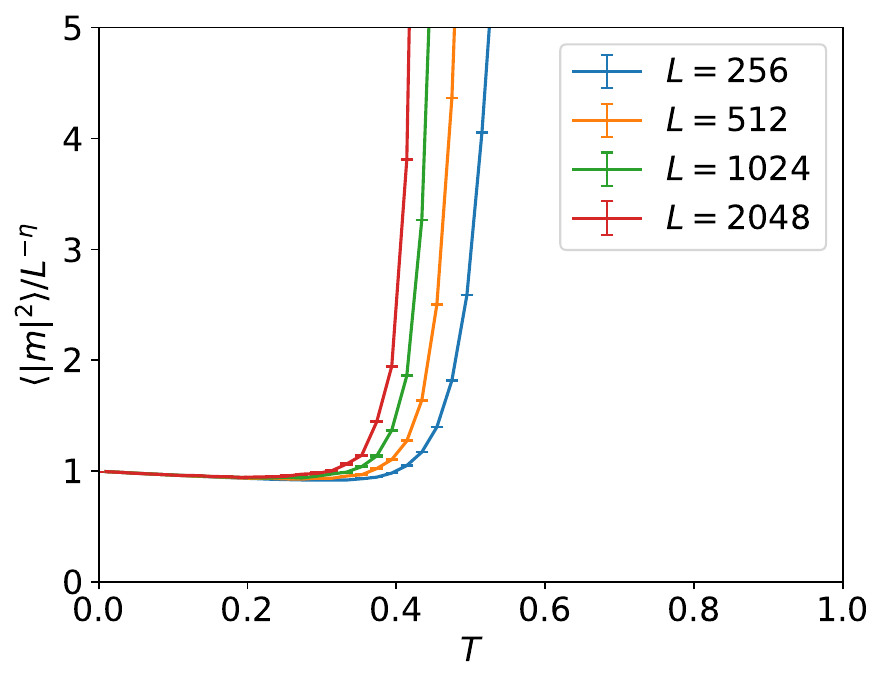} \\
        (\textbf{a})
    \end{minipage}
    \begin{minipage}{0.49\linewidth}
        \centering
        \includegraphics[width=0.99\linewidth]{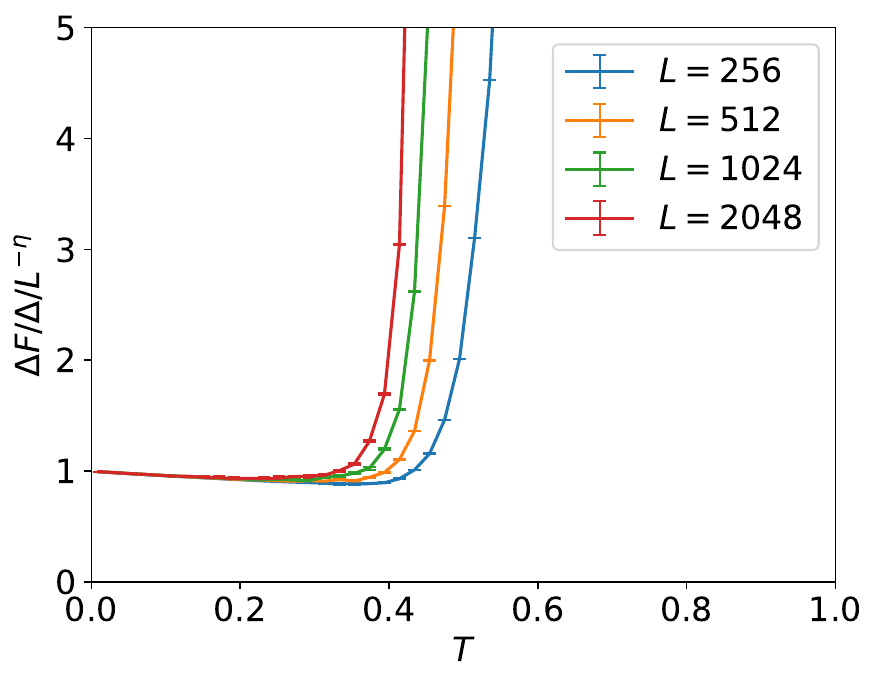} \\
        (\textbf{b})
    \end{minipage}
    \caption{
    Finite-size scaling of (\textbf{a}) the squared order parameter and (\textbf{b}) the cusp twist response.
    The data are rescaled as $\langle |m|^2\rangle L^{\eta(T)}$ and $Y_{\rm cusp}(L,T)L^{\eta(T)}$, where $\eta(T)$ is estimated from the correlation ratio.
    The low-temperature collapse confirms $\langle |m|^2\rangle\sim L^{-\eta(T)}$ and $Y_{\rm cusp}(L,T)\sim L^{-\eta(T)}$.
    }
    \label{fig:m2_eta_scaling}
\end{figure}

Figure~\ref{fig:m2_eta_scaling}a shows that the rescaled squared order parameter collapses well in the low-temperature regime.
This confirms that the zero-mode fluctuation follows the algebraic scaling expected from the correlation function.
Consequently, the order parameter vanishes in the thermodynamic limit for any finite temperature in the algebraically ordered phase.

We perform the same scaling analysis for the cusp twist response.
As discussed above, this quantity is the natural finite-size response associated with the nonanalytic spectral twist, but it does not behave as an independent stiffness analogous to the helicity modulus of the two-dimensional XY model.

Figure~\ref{fig:m2_eta_scaling}b shows that $Y_{\rm cusp}(L,T)L^{\eta(T)}$ also collapses in the low-temperature regime.
This indicates that the cusp response is governed by the same leading finite-size scaling as the squared zero-mode order parameter.
The collapse is lost in the crossover region, as expected because the low-temperature algebraic scaling no longer applies.

This result has an important implication.
In the conventional two-dimensional XY model, the helicity modulus remains finite throughout the low-temperature BKT phase, even though the order parameter vanishes in the thermodynamic limit.
In contrast, in the present fractional one-dimensional model, the cusp twist response scales in the same way as the squared zero-mode order parameter, $Y_{\rm cusp}(L,T)\sim \langle |m|^2\rangle\sim L^{-\eta(T)}$, and therefore vanishes in the thermodynamic limit for any finite temperature in the algebraically ordered phase.
Thus, the cusp twist response is not an independent stiffness that characterizes the BKT-like transition through a universal jump.
Rather, it is a finite-size observable governed by the same algebraic zero-mode scaling as $\langle |m|^2\rangle$.
The transition in the present model should therefore be characterized primarily by the change in the long-distance correlation function, the correlation ratio, the Binder ratio, and the effective exponent, rather than by a helicity-modulus jump.

\section{Discussion}\label{sec5}

\subsection{Nature of the Observed Transition}

The numerical results presented above indicate that the one-dimensional fractional $U(1)$ model with $\sigma=1$ exhibits a BKT-like transition.
The most direct evidence is obtained from the long-distance behavior of the correlation function and from dimensionless ratios constructed from it.
At low temperatures, the correlation function is consistent with a slow algebraic decay with a temperature-dependent exponent.
In this regime, the correlation ratio and the Binder ratio become approximately independent of system size, indicating a quasi-long-range ordered phase.

The low-temperature size-independent branches of $R_C$, the Binder ratio, and the effective exponent start to break down around $T\simeq0.35\text{--}0.4$.
Using $T_{\rm BKT}\simeq0.35$ as an operational estimate, the data for $R_C$ and the Binder ratio collapse well when plotted as functions of $(T-T_{\rm BKT})(\log L)^2$.
This scaling variable is expected when the relevant crossover length has a BKT-type essential singularity.
Thus, the finite-size scaling of the dimensionless ratios supports a BKT-type interpretation.

At the same time, the present transition is not identical to the conventional BKT transition of the two-dimensional XY model.
In the standard two-dimensional case, the high-temperature phase is characterized by exponential decay of correlations, whereas the low-temperature phase has algebraic correlations~\cite{Berezinskii1971,KosterlitzThouless1973,Kosterlitz1974}.
In the present nonlocal model, the disordered phase itself can contain an algebraic long-distance tail induced by the fractional kernel.
For $\sigma=1$, the Gaussian argument predicts an asymptotic high-temperature tail $C(r)\sim r^{-2}$, and this tail is observed numerically through the tendency of the effective exponent to approach $2$.
Therefore, the transition should not be described simply as a change from exponential to algebraic decay.
Rather, it is better interpreted as the emergence of a slower algebraic branch that dominates over the background $r^{-2}$ tail below the transition.

\subsection{Comparison with the Two-Dimensional XY Model}

The present model shares several features with the two-dimensional XY model.
True long-range order is absent at finite temperature in the marginal case $\sigma=1$, the low-temperature phase exhibits algebraic correlations with a continuously varying exponent, and the finite-size scaling of dimensionless ratios is consistent with an essential singularity of the crossover length.

However, the analogy with the two-dimensional BKT transition has a clear limitation.
In the two-dimensional XY model, the helicity modulus remains finite in the low-temperature phase and exhibits a universal jump at the transition~\cite{NelsonKosterlitz1977,WeberMinnhagen1988,HaradaKawashima1997}.
In the present one-dimensional fractional model, none of the twist responses examined here behaves as such a finite stiffness.
The ordinary linear-response quantity obtained by assuming an analytic $\delta^2$ twist cost grows with system size in the low-temperature regime, reflecting the nonanalyticity of the fractional dispersion.
The spectral cusp twist response is well defined at finite $L$ and has the correct zero-temperature normalization, but in the slow low-temperature algebraic branch it scales as $Y_{\rm cusp}(L,T)\sim L^{-\eta(T)}$, the same leading scaling as $\langle |m|^2\rangle$.
It therefore vanishes in the thermodynamic limit throughout the algebraically ordered phase.

Finite-twist responses also require care in the conventional two-dimensional XY model.
Khairnar and Vojta~\cite{KhairnarGauravVojtaThomas2025} showed that, for sufficiently large imposed twists, the
free-energy difference deviates from a simple quadratic dependence because mixing states of opposite chirality
produces an additional entropy contribution.
This effect is particularly relevant near antiperiodic boundary conditions and does not imply the absence of the conventional infinitesimal-twist stiffness.

The origin of the cusp in our spectral prescription is different: for $\sigma=1$, the uniform-configuration twist cost is already proportional to $|\Delta|$, as shown in Equation~\eqref{eq:uniform_twist_cost}.
Our comparison with the zero-mode fluctuations therefore addresses a different mechanism from the finite-angle
chirality-mixing effect in the local two-dimensional model.

This result shows that the cusp twist response is not an independent stiffness.
Rather, it probes the same algebraic finite-size scaling as the zero Fourier mode.
Consequently, the BKT-like transition in the present model cannot be characterized by a universal jump of a helicity modulus.
Instead, it must be characterized through correlation functions, correlation ratios, Binder ratios, and effective exponents.

\subsection{Role of the Nonlocal Tail}

A notable feature of the present fractional model is the coexistence of two different algebraic behaviors.
In the high-temperature regime, the nonanalytic Fourier multiplier $|k|^\sigma$ generates a long-distance tail $C(r)\sim r^{-(1+\sigma)}$.
For $\sigma=1$, this gives the $r^{-2}$ tail discussed above.
This algebraic decay is not a signature of quasi-long-range order; it is a background tail caused by the nonlocal kernel.

In contrast, below $T_{\rm BKT}$, the system develops a slower algebraic correlation with exponent smaller than $2$.
The correlation ratio and effective exponent distinguish these two regimes: in the low-temperature phase, $R_C(L,T)$ becomes size independent with $R_C\to2^{-\eta(T)}$, whereas in the high-temperature regime, once the $r^{-2}$ tail is reached, the same ratio tends to $1/4$.
Between these two regimes, the correlation ratio shows strong finite-size dependence.
This intermediate region corresponds to the crossover between the low-temperature algebraic branch and the high-temperature nonlocal tail.

This structure explains why the high-temperature side is more subtle than in short-range systems.
The disordered phase is not purely exponential at asymptotically long distances.
Instead, a large crossover length controls the approach to the $r^{-2}$ tail.
Near the transition, this crossover length becomes very large, and its finite-size scaling is consistent with a BKT-type essential singularity.

\subsection{Heuristic RG Interpretation}

The special role of $\sigma=1$ can be understood heuristically from a compact phase-only description {\color{black} at the Gaussian level}.
In the low-temperature regime where amplitude fluctuations are suppressed, one may write
\begin{align}
    q(x)\simeq e^{i\theta(x)}.
\end{align}

The Gaussian part of the phase Hamiltonian is
\begin{align}
    H_{\rm sw}
    = \frac{\rho_{\rm s}}{2} \sum_{k\neq0} |k|^\sigma |\theta_k|^2.
    \label{eq:RG_spin_wave}
\end{align}

{\color{black}Here $\rho_{\rm s}$ denotes the coefficient of the quadratic phase theory and is not identified a priori with the cusp twist response.}
Under a scale transformation $x\to e^\ell x$, the Gaussian coupling is marginal at $\sigma=1$.
For $\sigma<1$, long-wavelength phase fluctuations are weaker and true long-range order is not excluded by the spin-wave approximation.
For $\sigma=1$, the phase fluctuation grows logarithmically, producing algebraic correlations.
{\color{black}
However, Gaussian marginality alone does not establish a finite-temperature line of critical fixed points: interactions beyond the quadratic phase approximation must also be considered.
}

The fixed-amplitude compact phase energy is given by Equation~\eqref{eq:long_range_phase_coupling}.
Expanding the cosine for small phase differences gives Equation~\eqref{eq:RG_spin_wave}, but this approximation need not describe configurations with large phase differences.
In particular, a $2\pi$ change of the phase branch on an interval leaves $q=e^{i\theta}$ unchanged.
Assigning an energy to this branch change using an unwrapped Gaussian phase field therefore does not establish the interaction energy of physical defects.
Nontrivial winding configurations and transitions between winding sectors require a separate treatment of the compact field and, where relevant, its amplitude cores.
We consequently do not infer a logarithmic phase-slip-pair interaction solely from the Fourier symbol $|k|$.

For comparison, the conventional two-dimensional BKT mechanism involves both spin-wave fluctuations and vortices~\cite{Kosterlitz1974,JoseKadanoffKirkpatrickNelson1977}.
In a standard normalization, the leading RG equations are
\begin{align}
    \begin{aligned}
    \frac{dK_{\rm 2D}^{-1}}{d\ell}
    =4\pi^3 y_{\rm 2D}^2+O(y_{\rm 2D}^4), \qquad
    \frac{dy_{\rm 2D}}{d\ell}
    =(2-\pi K_{\rm 2D})y_{\rm 2D}+O(y_{\rm 2D}^3),
    \end{aligned}
    \label{eq:RG_standard_BKT_comparison}
\end{align}
where $K_{\rm 2D}=\rho_{\rm 2D}/T$ is the dimensionless coefficient
of the local phase-gradient energy, $\ell$ increases toward larger
length scales, and $y_{\rm 2D}$ is the vortex fugacity.
The fugacity is a statistical weight for inserting a vortex,
controlled in part by its core energy, rather than the vortex
density itself.
The first equation describes the screening of the stiffness by
vortex pairs, while the second determines whether vortex fugacity
grows or decreases.
Near $\pi K_{\rm 2D}=2$, introducing
$x_{\rm KT}=\pi K_{\rm 2D}-2$ and
$z_{\rm KT}=4\pi y_{\rm 2D}$ reduces the leading flow to
\begin{align}
    \frac{dx_{\rm KT}}{d\ell}=-z_{\rm KT}^2,
    \qquad
    \frac{dz_{\rm KT}}{d\ell}=-x_{\rm KT}z_{\rm KT},
    \label{eq:RG_KT_normal_form_comparison}
\end{align}
up to higher-order terms.
For a generic thermal trajectory crossing the separatrix, this flow~yields
\vspace{-6pt}\begin{align}
\log[\xi_{\rm 2D}(T)/a]\sim
\frac{b_{\rm 2D}}{\sqrt{T-T_{\rm BKT}^{\rm(2D)}}}
\end{align}
on the disordered side, where $a$ is a microscopic length and
$b_{\rm 2D}>0$ is nonuniversal.
This explains the essential singularity underlying the standard
BKT finite-size scaling form.

Equations~\eqref{eq:RG_standard_BKT_comparison}
and \eqref{eq:RG_KT_normal_form_comparison} are presented as a
comparison, not as RG equations derived for the present fractional
model.
A corresponding phase-slip fugacity and its flow would require a
physically defined defect ensemble and a controlled coarse-graining
of the original nonlocal energy.
In particular, the defect-induced renormalization of the long-range
coupling cannot be assumed to have the same form as vortex
screening in two dimensions.
Phase-slip events along the stochastic sampling trajectory do not,
by themselves, define a static defect gas in the one-dimensional
equilibrium measure.

Related studies of two-dimensional long-range models emphasize the need for a model-specific treatment of topological excitations.
Giachetti et al.~\cite{Giachetti2023} investigated a long-range generalization of the Villain model and found that its critical behavior need not coincide with that of the corresponding long-range XY model.
Their analysis highlights the limitations of assuming that spin waves and topological defects can be decoupled in the presence of long-range couplings.

More recently, Walther et al.~\cite{WaltherLuisWillsherJosefKnolleJohannes2026} included the coupling between spin-wave and vortex excitations in their analysis of the two-dimensional long-range XY model.
They argued that spin-wave renormalization of the vortex interaction stabilizes the BKT transition.
These results concern vortices in two spatial dimensions and do not directly establish a phase-slip fugacity flow for the present one-dimensional model.
They nevertheless reinforce the need to derive the defect description and its coupling to phase fluctuations from the specific nonlocal Hamiltonian.

A relevant alternative is provided by the low-temperature RG analysis of the fixed-length inverse-square $O(n)$ chain by Hofstetter and Zwerger~\cite{Hofstetter1998}.
Writing $\vartheta$ for their dimensionless running temperature and $n$ for the number of spin components, their normalization gives, in one spatial dimension,
\begin{align}
    \frac{d\vartheta}{d\ell}
    =(n-1)\vartheta^2+O(\vartheta^3).
    \label{eq:RG_inverse_square_comparison}
\end{align}

For $n=2$, the positive leading term does not vanish.
Their RG describes zero-temperature criticality, with a characteristic length growing exponentially in inverse
temperature, rather than a finite-temperature BKT fixed line.
Thus, an essential growth of a characteristic length does not by itself identify the finite-temperature BKT mechanism.
The fixed-amplitude limit of the present model has the same leading $|k|$ kernel, making this comparison directly relevant.
Allowing amplitude fluctuations or changing the ultraviolet regularization does not, by itself, establish a different infrared scenario.

Accordingly, the Gaussian analysis motivates the study of $\sigma=1$, but it does not provide a microscopic derivation of phase-slip unbinding.
The slow correlations and approximate BKT-type collapse reported above characterize the accessible finite-size regime; by themselves, they do not exclude a long crossover associated with zero-temperature criticality.
Here the term BKT-like therefore refers to the observed correlation and finite-size scaling phenomenology, while the estimated $T_{\rm BKT}$ remains an operational parameter of that analysis.
Establishing a genuine finite-temperature transition, or distinguishing it from an extended crossover, requires further control of the long-distance limit and of the RG of the compact nonlocal model.

It is worth noting that BKT-like stiffness--fugacity flows have been derived in other nonstandard settings, for example in the dual description of compact $(2+1)$-dimensional Abelian Higgs models, where anomalous gauge-field scaling produces logarithmic defect interactions and an anomalous $k^3$ gradient term
~\cite{KleinertHagenNogueiraFlavioSudboAsle2002,HagenKleinertFlavioNogueiraAsleSudbo2003}.
This provides an instructive comparison, but does not by itself establish the corresponding defect ensemble or fugacity flow for the present one-dimensional compact fractional model.

\section{Summary and Outlook}\label{sec6}

We studied a one-dimensional fractional $U(1)$ Ginzburg--Landau model with quadratic dispersion $|k|^\sigma$, focusing mainly on the marginal case $\sigma=1$.
The model is nonlocal in real space, with an interaction kernel decaying as $1/|x-y|^{1+\sigma}$.
For $\sigma=1$, the kernel decays as $1/|x-y|^2$, and the spin-wave approximation predicts logarithmic phase fluctuations.

\textls[-15]{Using stochastic Gross--Pitaevskii dynamics as an equilibrium sampler of the Gibbs measure, we investigated the correlation function, correlation ratio, Binder ratio, effective exponent, and twist responses.
Our main findings are as follows.
First, the low-temperature phase exhibits algebraic correlations with a temperature-dependent exponent.
In this regime, the correlation ratio, Binder ratio, and effective exponent are nearly independent of system size.
Second, the low-temperature branch loses its size-independent character around $T\simeq0.35\text{--}0.4$, and the finite-size scaling analysis is consistent with $T_{\rm BKT}\simeq0.35$.
With this operational estimate, the correlation ratio and Binder ratio show good collapse as functions of \mbox{$(T-T_{\rm BKT})(\log L)^2$}, supporting a BKT-type essential scaling of the crossover length.}

Third, the high-temperature regime is not characterized by a simple exponential correlation decay at asymptotically long distances.
Because of the nonanalytic fractional kernel, the disordered regime contains a long-distance algebraic tail.
For $\sigma=1$, this tail behaves as $C(r)\sim r^{-2}$.
Thus, the transition is better understood as the emergence of a slower algebraic branch with exponent smaller than $2$, rather than as a simple change from exponential to algebraic decay.

Fourth, twist-response quantities behave differently from the helicity modulus in the two-dimensional XY model.
The ordinary linear-response helicity-modulus-like quantity grows with system size and is not a finite stiffness.
The cusp twist response is finite at each system size, but in the slow low-temperature algebraic branch it scales as \mbox{$Y_{\rm cusp}(L,T)\sim L^{-\eta(T)}$}, the same leading scaling as $\langle |m|^2\rangle$.
Therefore, the cusp response vanishes in the thermodynamic limit and cannot serve as an analogue of the helicity modulus of the two-dimensional XY model.

These results indicate that the $\sigma=1$ one-dimensional fractional $U(1)$ model realizes a BKT-like transition in terms of correlation functions and logarithmic finite-size scaling, but not in terms of a helicity-modulus jump.
The transition is therefore similar to the two-dimensional BKT transition in its algebraic low-temperature phase and essential finite-size scaling, but different in its nonlocal high-temperature tail and in the absence of an independent stiffness.

Several open problems remain.
{\color{black} A more systematic RG analysis of the compact fractional phase model would help clarify the origin of the observed BKT-like finite-size scaling and distinguish a genuine finite-temperature transition from an extended crossover.
Such an analysis should establish whether a defect-based description is appropriate and, if so, how the corresponding fugacity couples to long-distance phase fluctuations.}
It would also be useful to study other values of $\sigma$, especially the crossover from the marginal case $\sigma=1$ to the regime $\sigma<1$, where true long-range order may appear.
Finally, the relation between fractional nonlocality, stochastic Gross--Pitaevskii dynamics, and possible physical realizations in long-range superfluid or condensate systems deserves further investigation.

\vspace{6pt} 

\section*{Funding}
This research was funded by JSPS KAKENHI Grants No. 23K22492, No. 24K00593, and No.~26K07020.

\section*{Data Availability Statement}
The processed numerical data used to generate the figures in this article, together with the simulation and analysis codes, are openly available at \url{https://www.kochi-tech.ac.jp/~michikaz/monkas/2026/1D_fractional_BKT/} (accessed on 7 Oct 2026).
Additional simulation outputs are available from the corresponding author upon reasonable request.

\bibliographystyle{unsrtnat}
\bibliography{paper_MDPI}

\end{document}